\documentclass[secnumarabic,aps,prb,reprint,superscriptaddress]{revtex4-2}
\usepackage[english]{babel}
\usepackage{graphicx}
\usepackage{color}
\usepackage{amsmath}
\usepackage{tabularx}
\usepackage[dvipsnames]{xcolor}

\begin{document}

\title{Exchange bias without directional anisotropy in Permalloy/CoO bilayers}
\author{Alexander Mitrofanov}
\affiliation{Department of Physics, Emory University, Atlanta, GA, USA.}
\author{Guanxiong Chen}
\affiliation{Department of Physics, Emory University, Atlanta, GA, USA.}
\author{Alexander Kozhanov}
\affiliation{Department of Electrical and Computer Engineering, Duke University, Durham, NC, USA.}
\author{Sergei Urazhdin}
\affiliation{Department of Physics, Emory University, Atlanta, GA, USA.}

\begin{abstract}

We utilize transverse ac susceptibility measurements to characterize magnetic anisotropy in archetypal exchange-bias bilayers of ferromagnet Permalloy (Py) and antiferromagnet CoO. Unidirectional anisotropy is observed for thin Py, but becomes negligible at larger Py thicknesses, even though the directional asymmetry of the magnetic hysteresis loop remains significant. Additional magnetoelectronic measurements, magneto-optical imaging, as well as micromagnetic simulations show that these surprising behaviors are likely associated with asymmetry of spin flop distribution created in CoO during Py magnetization reversal, which facilitates the rotation of the latter back into its field-cooled direction. Our findings suggest new possibilities for efficient realization of multistable nanomagnetic systems for neuromorphic applications.

\end{abstract}

\maketitle

\section{Introduction}

Intense research in nanomagnetism has facilitated dramatic improvements in  information storage and data processing technologies, owing to ever-increasing memory densities, read and write speeds. In hard disks or magnetic random access memory, information is stored in the magnetization state of bistable nanomagnets~\cite{piramanayagam2012developments}, where magnetocrystalline, interfacial, and/or shape anisotropies produce a large magnetic energy barrier between stable magnetization states. Long-term stability requires that this barrier significantly exceeds thermal energy, $k_BT=25$~meV for devices operating at room temperature (RT) $T=295$~K. Here $k_B$ is the Boltzmann constant.

Advances in materials science have enabled effective anisotropy fields exceeding $1$~Tesla, resulting in anisotropy barriers approaching $100$~$\mu$eV per magnetic atom~\cite{1396176,TUDU2017329}. This energy scale sets the ultimate limit for the minimal size of magnetically stable volume to about $10^4$ atoms. In practice, the anisotropy barrier is lowered by imperfections and inhomogeneous magnetization states formed during magnetization reversal~\cite{PhysRevB.96.014412}.

Even for the largest achievable magnetic anisotropies, the associated energy density is three orders of magnitude smaller than the exchange energy, about $100$~meV per atom for transition-metal ferromagnets~\cite{Nanayakkara2016,}. If exchange could be utilized to stabilize the magnetization state, 10 magnetic atoms would be sufficient to store a bit of information. Such a small volume is impractical for magnetic or magnetoelectronic readout of individual magnetic bits. Nevertheless, exchange-dominated large-scale magnetic energy landscape may stabilize a multitude of magnetic configurations in a nanoscale volume, facilitating efficient implementation of non-binary neuromorphic nanodevices such as magnetic memristors~\cite{4781542,PhysRevB.78.113309,Krzysteczko2009,Krzysteczko2012}.

Ultrasmall multistable nanomagnetic systems based on exchange interaction can be developed by taking advantage of magnetic frustration. Indeed, the ground state of geometrically frustrated spin systems is massively degenerate~\cite{Balents2010}. However, geometrically frustrated spin systems rely on a subtle balance of structural and electronic effects, which may be too delicate for robust room-temperature device applications. In randomly frustrated magnetic systems such as spin glasses, random exchange interactions produce a hierarchical energy landscape, resulting in a multitude of Gibbs states~\cite{RevModPhys.58.801}. Unfortunately, conventional dilute spin glasses are not amenable to efficient magnetoelectronic control and detection of magnetic state required for the memristor operation.

Thin-film ferromagnet/antifferomagnet (F/AF) heterostructures may provide an alternative to spin glasses that can combine frustration effects with efficient control and detection of the magnetization state~\cite{doi:10.1063/5.0018411}. Indeed, F layers in such heterostructures are amenable to magnetoelectronic readout and can be manipulated by spin torque~\cite{PhysRevApplied.12.044029}, while the frustration of exchange interaction at the F/AF interface, generally expected due to the incompatibility between F and AF magnetic orderings~\cite{PhysRevB.35.3679,PhysRevB.101.144427}, can result in glassy behaviors stabilizing a multitude of magnetization states~\cite{PhysRevApplied.12.044029,PhysRevB.92.174416,PhysRevB.97.054402}. Furthermore, the effects of frustrated exchange interaction in such systems can be tailored by the magnetic properties and thicknesses of individual layers, and by the strength of their coupling~\cite{PhysRevB.94.024422}.

F/AF bilayers have been extensively studied in the context of exchange bias (EB) - asymmetry of the hysteresis loop observed upon field-cooling~\cite{PhysRev.105.904,NOGUES1999203, KIWI2001584}. The EB effect is generally understood as a consequence of the formation of stable uncompensated AF magnetic moments at the F/AF interface. Their exchange interaction with the magnetization $\mathbf{M}$ of F can be described as an effective exchange field $H_E$ shifting the hysteresis loop of the latter. The ''pinning" of $\mathbf{M}$ due to EB has found extensive applications in magnetic recording and sensing~\cite{Nogus2005}.

Neuromorphic applications of F/AF heterostructures require the ability to stabilize multiple magnetization configurations and efficiently switch among them. However, a generally accepted picture of the microscopic spin state in these systems, necessary for its efficient manipulation, has not yet emerged. Instead, numerous competing models have been developed over more than 60 years of relevant research, as discussed in multiple reviews~\cite{NOGUES1999203, Stamps_2000, KIWI2001584,Nogus2005,BERKOWITZ1999552,Giri_2011}. Below, we outline only the models that we believe are most directly relevant to our findings.

Mauri et al.~\cite{Mauri1987} proposed that the reversal of the magnetization $\mathbf{M}$ of F in F/AF bilayers results in ''winding" of an exchange spring in AF, i.e., twisting of the N\`eel order through the AF thickness due to the exchange coupling at their interface. The difference in energy between the ''wound" and the ''unwound" state can be interpreted as the effective Zeeman energy associated with the exchange-bias field $H_E$. This model was supported by the experimental observations of partial rotation of AF spins due to the reversal of $\mathbf{M}$~\cite{PhysRevLett.91.017203,PhysRevLett.92.247201}. 

In a qualitatively different approach, Malozemoff suggested that frustrated exchange interaction at F/AF interfaces can be described as a random effective exchange field acting on AF~\cite{PhysRevB.35.3679}, which breaks AF up into Imry-Ma domains~\cite{PhysRevLett.35.1399}. Conversely, frustrated F/AF exchange should also manifest as effective random fields acting on F, which has been experimentally confirmed~\cite{Rezende,PhysRevB.101.144427,jimenez2009emergence}. For small AF thicknesses, Malozemoff predicted a crossover to the ''Heisenberg domain state" (HDS), in which the AF domain sizes become smaller than the domain wall (DW) widths, so the N\`eel order becomes twisted everywhere~\cite{PhysRevB.37.7673}. Such a crossover was recently experimentally observed~\cite{PhysRevB.94.024422}, and HDS was identified as a correlated spin glass state - a state that exhibits short-range AF spin correlations but lacks long-range ordering - formed below a certain material-dependent thickness of AF~\cite{PhysRevB.97.054402,urazhdin2019JMMM}.

In yet another theoretical insight, Koon showed that the exchange energy at F/AF interface is minimized by the AF spin flop - rotation of the N\`eel order into direction orthogonal to the $\mathbf{M}$, accompanied by tilting of the AF magnetic sublattices into or opposite to the direction of $\mathbf{M}$, depending on the sign of F/AF exchange~\cite{PhysRevLett.78.4865}. Schulthess and Butler pointed out that according to this picture, reversal of $\mathbf{M}$ should result in the reversal of the AF spin tilting, leading to uniaxial rather than unidirectional anisotropy~\cite{PhysRevLett.81.4516}. Indeed, a large uniaxial anisotropy is commonly observed in F/AF bilayers~\cite{NOGUES1999203,BERKOWITZ1999552}.

It is currently believed that different combinations of the proposed mechanisms of EB are likely responsible for the wide variety of experimental observations for different F/AF systems. However, to the best of our knowledge, microscopic mechanisms most relevant to specific materials or systems have not yet been identified.

Here, we present experimental results that allow us to tentatively identify the mechanisms governing magnetization states in a ''classic" extensively studied EB system comprising antiferromagnet CoO and a model low-anisotropy ferromagnet Permalloy (Py). Our main result is highly counter-intuitive: unidirectional anisotropy is observed for thin Py, but vanishes for thicker Py, even though the asymmetry of the hysteresis loop persists. We use several additional complementary characterization techniques and micromagnetic simulations to show that our results are in fact simultaneously consistent with all the theoretical models outlined above, with one caveat - Mauri's exchange spring-like behaviors are likely associated with magnetic history-dependent distribution of local AF moments rather than coherent N\'eel order twisting. Our findings provide microscopic insight into frustrated magnetism that may become crucial for the implementation of nanometer-scale neuromorphic nanodevices based on F/AF heterostructures.

The paper is organized as follows. In the next section, we provide details on sample fabrication and our experimental approach. In Section III, we present our experimental results. In Section IV, we introduce a microscopic model of the studied system, and present the results of  micromagnetic simulations based on this model, which capture the salient features of our experimental observations. Section V summarizes our findings.

\section{Materials and experimental methods}\label{sec:methods}

\subsection{Sample preparation}

CoO(8)Py(t)Ta(2) magnetic multilayers with Permalloy (Py) thickness $t$ varied between $20$ and $50$~nm were grown on silicon substrates by sputtering at room temperature, at the base pressure of $5\times10^{-9}$~Torr. Numbers in parenthesis are thicknesses in nanometers. Ta(2) served as a capping layer protecting the films from oxidation. The thickness of $8$~nm for CoO was above the transition to the Heisenberg domain state (HDS) at $t\approx 6$~nm, minimizing the glassy magnetic dynamics of the AF magnetic order driven by the Py magnetization~\cite{PhysRevB.94.024422,urazhdin2019JMMM}. Nevertheless, our data provide evidence of spin flop reversal in AF, which is expected even for large AF thickness~\cite{PhysRevLett.81.4516}.

The multilayers were deposited in a $150$~Oe in-plane magnetic field, which is known to facilitate magnetic ordering in CoO. Py and Ta were deposited by dc sputtering from the stoichiometric targets in $1.8$~mTorr of ultrapure Ar, while CoO was deposited from a Co target by reactive sputtering in a mixture of ultrapure oxygen and Ar, with the partial pressure of oxygen optimized as in our previous studies of CoO-based systems~\cite{PhysRevB.78.052403,PhysRevB.94.024422,urazhdin2019JMMM, PhysRevB.101.144427}. Measurements  below the N\`eel temperature $T_N=290$~K of CoO, were performed after cooling from room temperature (RT) in the presence of an external magnetic field $H_{dc}=0.5$~kOe.

\subsection{Transverse ac susceptibility}

\begin{figure}
	\includegraphics[width=0.9\columnwidth]{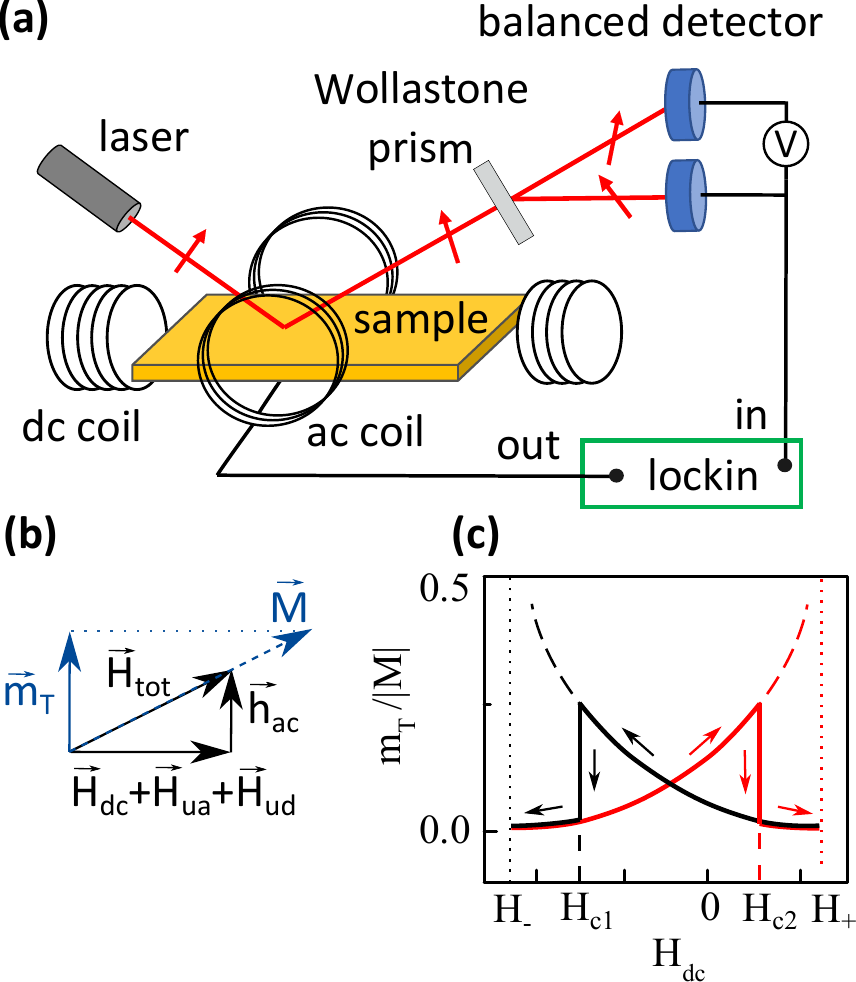}	
	\caption{\label{fig:meas_schematic} (Color online) (a) Schematic of the MOKE setup combining transverse ac susceptibility and longitudinal dc magnetometry. (b) Effective magnetic fields that determine the instantaneous orientation of the magnetization $\mathbf{M}$ in the quasi-static limit, when the frequency of the ac field $\mathbf{h}_{ac}$ is far below FMR. (c) Calculated ac susceptibility vs dc field, in the presence of uniaxial and unidirectional anisotropies. The dotted vertical lines show the fields at which the  extrapolated linear susceptibility diverges. The dashed vertical lines show the coercive fields.}
\end{figure}

The transverse ac susceptibility technique is utilized to precisely characterize magnetic anisotropy, by detecting oscillations of magnetization in response to ac field $\mathbf{h}_{ac}(t)=\sqrt{2}\mathbf{H}_{ac}\cos\omega t$, with a saturating dc field  $\mathbf{H_{dc}}$ applied perpendicular to it~\cite{PhysRevB.62.8931,PhysRevB.68.220401,Rezende2003,PhysRevB.76.054409,urazhdin2019JMMM}. All the results presented below were obtained with $H_{ac} =8$~Oe rms at frequency $\omega/2\pi=1.3$~kHz. Our ac susceptibility measurement setup, schematically shown in~Fig.~\ref{fig:meas_schematic}(a), utilizes magneto-optical Kerr effect (MOKE) to detect the oscillations of $\mathbf{M}$ driven by $\mathbf{h}_{ac}$. The component $M_\parallel$ of $\mathbf{M}$ along the dc field is also measured, allowing us to simultaneously obtain the usual longitudinal magnetic hysteresis loop. The sample is placed on the cold finger fabricated from undoped Si to minimize inductive shunting of ac field. All the fields are in-plane, with $\mathbf{H}_{dc}$ produced by an external electromagnet, and $\mathbf{h}_{ac}$ generated by a coil built into the cold finger to minimize artifacts due to induction.

A p-polarized laser beam with the spot size of about $2$~mm is incident on the sample at an angle of $45^\circ$ to the film surface, with $H_{dc}$ in the incidence plane, and $\mathbf{h}_{ac}(t)$ perpendicular to it. The polarization of the beam reflected by the magnetic film is rotated due to the longitudinal MOKE effect, with the rotation angle proportional to $M_\parallel$~\cite{Qiu2000}. The Kerr polarization rotation is detected using a balanced photodetector scheme, with a Wollastone prism used to split the beam into two orthogonally polarized beams. The intensity of the beam reflected by the sample also oscillates due to transversal MOKE associated with the oscillating in-plane magnetization component $m_\perp(t)$ perpendicular to the incidence plane. This intensity oscillation is detected by the lock-in amplifier connected to the output of one of the photodetectors. Thus, both dc and ac measurements can be performed simultaneously in this geometry.

The measurement geometry of transverse susceptibility is the same as in the ferromagnetic resonance (FMR) technique, which utilizes a microwave field perpendicular to the dc field to resonantly drive magnetization oscillations. In ac susceptibility measurements, the frequency of the ac field is typically in the Hz to kHz range, far below the FMR resonance. In this quasi-static limit, the resonant dynamical properties of F are irrelevant. Its magnetization $\mathbf{M}$ simply follows the net effective magnetic field comprising the external field and the effective anisotropy field, allowing one to precisely characterize the latter, as follows.

Consider an F/AF bilayer cooled in field $H_{cool}>0$, setting the directions of the effective exchange bias field $H_{ud}$ and the effective uniaxial anisotropy field $H_{ua}$ accompanying EB. We define these fields by the corresponding contributions to the total magnetic energy of F per unit volume averaged through its thickness, $E_M=-M[(H_{dc}+H_{ud})\cos\theta+H_{ua}\cos^2\theta/2]$, where $\theta$ is the angle between the magnetization and the effective fields. 

The magnetization $\mathbf{M}$ of F oscillates in-phase with the ac field $h_{ac}(t)$, following the direction of the total field $\mathbf{H_{eff}+h_{ac}(t)}$, as shown in Fig.~\ref{fig:meas_schematic}(b). We have verified that the out-of-phase ac component of $M_\perp(t)$ for the studied samples was negligible compared to the in-phase component. In the limit $h_{ac}\ll H_{eff}$ applicable to our measurements, $m_T(t)\approx |M|h_{ac}(t)/H_{eff}=\sqrt{2}M_T\cos\omega t$, where 

\begin{equation}\label{eq:M_T}
\frac{M_T}{|M|}=\frac{H_{ac}}{|H_{eff}|}=\frac{H_{ac}}{|H_{dc}|+H_{ua}\pm H_{ud}}
\end{equation}

is the rms ac magnetization component normalized by the total magnetization. The ''+" (''-") sign in the denominator corresponds to saturated state at $H_{dc}>0$ ($<0$). The transverse ac magnetic susceptibility $\chi_T$ was defined in some prior studies as the ratio of the amplitude of the ac magnetization component to the amplitude of the ac field~\cite{PhysRevB.62.8931,Rezende2003,PhysRevB.76.054409,urazhdin2019JMMM}. However, in contrast to the linear susceptibility of demagnetized systems measured in the absence of dc fields, this quantity by itself is not fundamentally significant, since it is proportional to the magnetization, and is dependent on the dc field. In contrast, response characterized by $M_T/M$ is dimensionless and directly reflects the geometric aspects of the measurement. Below, we colloquially refer to this quantity, as well as $M_T$, as the transverse susceptibility. 

The dependence of $M_T/M$ on $H_{dc}$ described by Eq.~(\ref{eq:M_T}) is schematically shown in Fig.~\ref{fig:meas_schematic}(c).
Starting at a large $H_{dc}>0$ along the cooling field direction, $M_T$ increases with decreasing $H_{dc}$, and may be expected to diverge at $H_{dc}\equiv H_-=-H_{ua}-H_{ud}$, which corresponds to the stability loss. In practice, the magnetization reverses at the coercive field $H_{C1}>H_-$.

In the reversed magnetization state, the effective uniaxial anisotropy field changes sign, while the effective unidirectional anisotropy field remains the same. Thus, as the dc field is increased from a large negative value, $M_T/M$ increases and is extrapolated to diverge at $H_+=H_{ua}-H_{ud}$, but the magnetization reverses at $H_{C2}<H_+$. Based on the above relations, the effective unidirectional and uniaxial anisotropy fields can be obtained from $H_-$ and $H_+$ determined by fitting the measured $M_T(H_{dc})/M$ with Eq.~(\ref{eq:M_T}), as $H_{ud}=(H_++H_-)/2$, $H_{ua}=(H_+-H_-)/2$. 

Some of the prior studies of transverse ac susceptibility analyzed the dependence $\chi^{-1}_T(H_{dc})$, or equivalently $M^{-1}_T(H_{dc})M$, the inverse of the quantity discussed here~\cite{PhysRevB.62.8931,Rezende2003,PhysRevB.76.054409}. According to Eq.~(\ref{eq:M_T}), these dependences are linear and extrapolate to the intercepts $H_+$, $H_-$, simplifying the analysis. However, this approach places larger weight on small ac signals at large $H_{dc}$, leading to larger errors in $H_+$, $H_-$.

In the studies of EB based on hysteresis loop measurements, a different set of characteristics termed EB field and coercivity are commonly introduced, defined as $H_{EB}=(H_{C2}+H_{C1})/2$ and $H_C=(H_{C2}-H_{C1})/2$, respectively. The main result of this paper is that the relation between these two sets of parameters describing EB is dependent on the thickness of F, revealing the microscopic mechanisms of EB in the studied system.

\subsection{Magneto-optical imaging} 

Magneto-optical microscopy~\cite{doi:10.1063/1.4991820, doi:10.1063/1.5003719} was utilized to verify that the effects observed in transverse susceptibility measurements cannot be attributed to some complex inhomogeneous magnetization states, and to elucidate the reversal mechanism. In our custom-built MOKE microscopy setup, light produced by a 430~nm light-emitting diode (LED) passes through a collimating system, which includes a rectangular diaphragm whose image is focused in the backplane of the 50x  objective with 0.85 numeric aperture, and is shifted from its axis to define the average incidence angle around $40^\circ$. A polarizing beamsplitter is used to polarize the incident light in the incidence plane, and to direct the reflected light to the detection path, which includes an analyzer, a lens, and a camera. Real-time imaging with temporal resolution of about 15~ms is limited by the sensitivity of the camera, and spatial resolution of about 250~nm is diffraction-limited. A two-stage Peltier cooling system enables variable temperature measurements, with the base temperature of $218$~K.  

\subsection{Magnetoelectronic characterization} 

Magnetoelectronic measurements based on the anisotropic magnetoresistance (AMR) of Py were utilized to independently confirm the anisotropy crossover observed from the susceptibility measurements. 
The main advantage of such measurements, compared to the traditional magnetometry, is their high sensitivity to the components of magnetization non-collinear with the magnetic field. For instance, measurements discussed below were sufficiently sensitive to detect average angles of about $2-3^\circ$ between the magnetization and the field. The corresponding deviations from the saturated magnetization value are less than $0.1\%$, beyond the typical detection limit of standard magnetometry set by the magnetic background.

The magnetoelectronic measurements were performed in the 4-probe van der Pauw geometry, with the current flowing in the direction of the long side of the $6$~mm$\times 2$~mm rectangular sample. ac current of $0.1$~mA rms at a frequency of $1.3$~kHz was applied to the sample, and the ac voltage was detected by the lockin amplifier. 

AMR is $180^\circ$-periodic with respect to the angle between the magnetization $\mathbf{M}$ and the direction of electrical current $I$ flowing through Py, and is minimized when they are orthogonal. In our measurements, the in-plane field was precisely aligned orthogonal to the direction of current, by utilizing the dependence of AMR on the field direction at RT. Alignment precision of about $1^\circ$ was limited by the resolution of the electronic measurement. In this geometry, the dependence of resistance on the angle $\theta$ between the field and magnetization $M$ of Py is described by $R(\theta)=R_0+\Delta R\sin^2(\theta)$, where $R_0$ is the resistance minimum at $\theta=0^\circ$, and $\Delta R$ is the  magnetoresistance.  In this geometry, the technique is sensitive to the component of $\mathbf{M}$ non-collinear with $\mathbf{H}_{dc}$, with the resistance increase proportional to average $\theta^2$.

\section{Experimental results} 

\subsection{Magnetic hysteresis and ac susceptibility}

\begin{figure}
	\includegraphics[width=0.85\columnwidth]{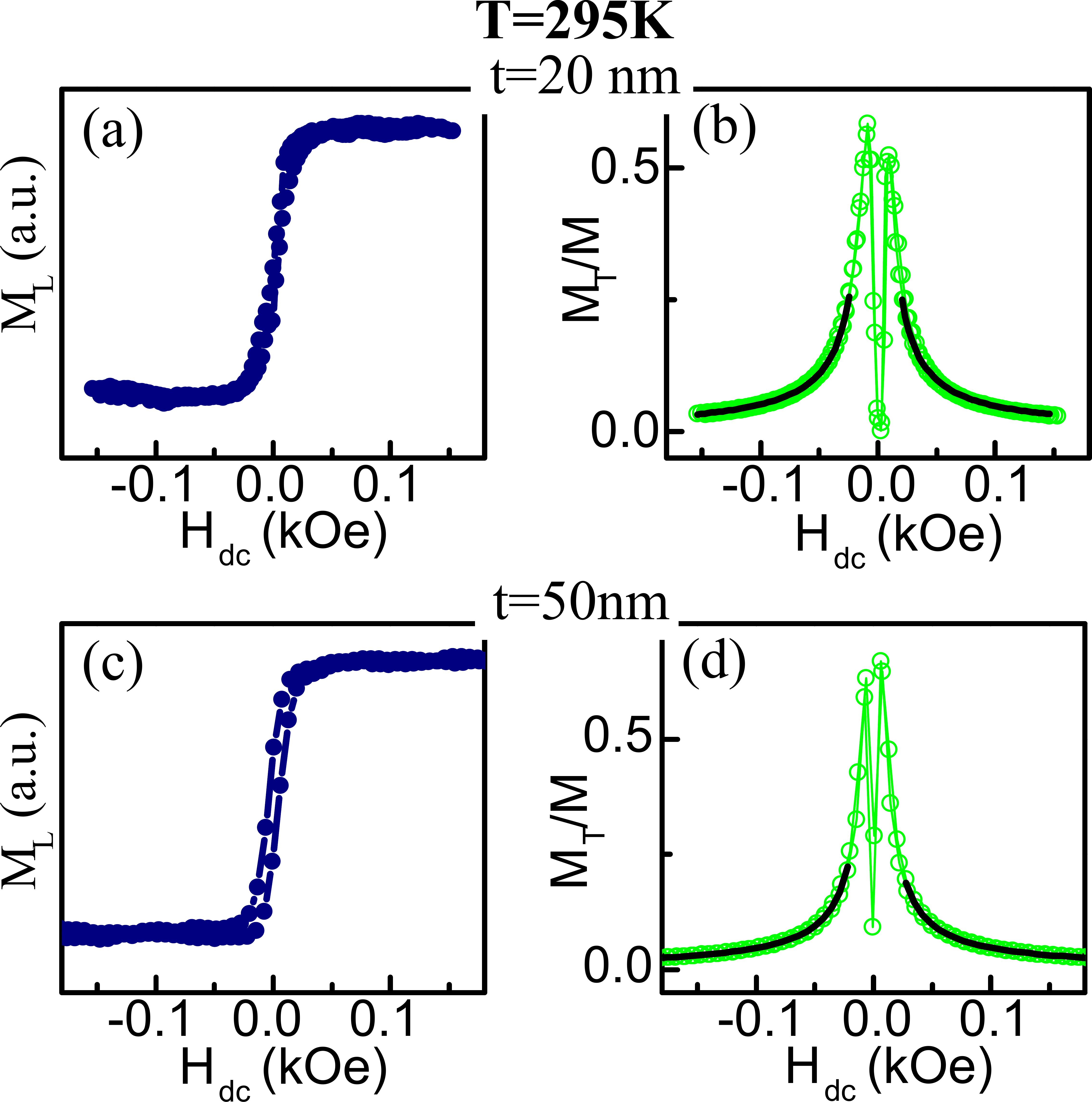}
	\caption{\label{fig:suscept_RT} (Color online) Longitudinal dc MOKE hysteresis loops (a,c) and transverse ac susceptibility (b,d) acquired at $T=295$~K for Py($t$)/CoO(8) bilayers with $t=20$~nm (a,b) and $50$~nm (c,d).}
\end{figure}

We start by analyzing the measurements at RT $T=295$~K above the Neel temperature $T_N=290$~K of CoO. In this case, the magnetic properties of Py/CoO are expected to be similar to those of standalone Py films. Indeed, the longitudinal dc MOKE hysteresis curves are consistent with the soft low-anisotropy magnetic properties of Py [Fig.~\ref{fig:suscept_RT}(a,c)]. The transverse ac susceptibility curves [Fig.~\ref{fig:suscept_RT}(b,d)] exhibited negligible hysteresis, sharply peaking at  $H_{dc}$ approaching $0$, as expected from Eq.~(\ref{eq:M_T}) for negligible anisotropy, i.e., $H_{eff}=H_{dc}$. Large dips at $H_{dc}=0$ are associated with the formation of a multidomain state during the magnetization reversal, resulting in vanishing susceptibility at ac fields below the domain wall depinning.

Fitting of the ac susceptibility data with Eq.~(\ref{eq:M_T}) [curves in Fig.~\ref{fig:suscept_RT}(b,d)] yielded negligible effective anisotropy fields. We used the scaling between $M_T/M$ and the measured ac magneto-optic signals inferred from the fitting for the analysis of low-temperature data. This reduced the number of fitting parameters to one (the anisotropy field), minimizing the fitting uncertainty. Separate tests for standalone Py films confirmed negligible temperature dependence of the magneto-optic coefficients.

\begin{figure}
	\includegraphics[width=0.7\columnwidth]{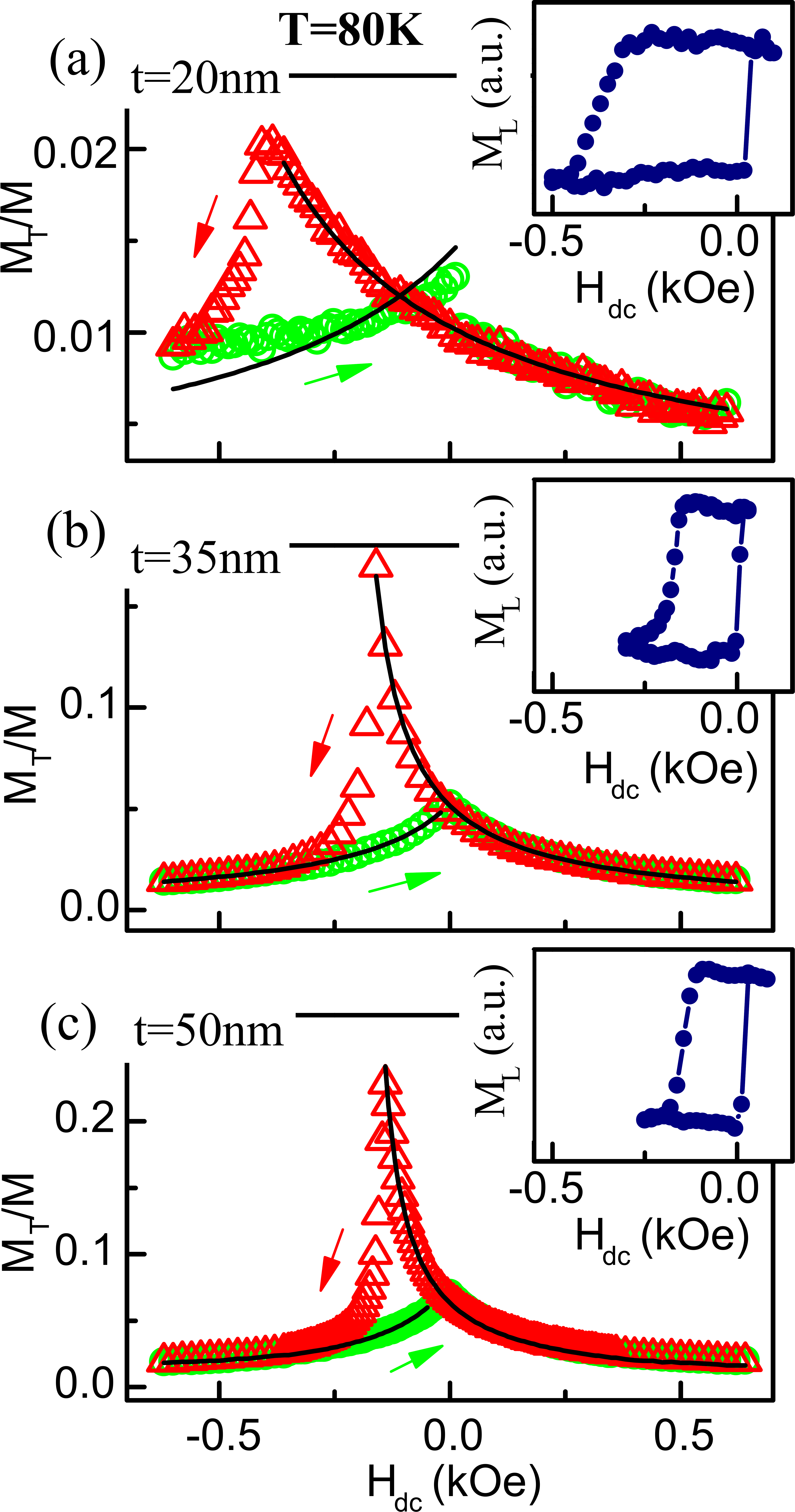}
	\caption{\label{fig:suscept_80K} (Color online) ac susceptibility vs dc field for Py($t$)/CoO(8) bilayers with Py thicknesses $t=20$~nm (a), $35$~nm (b), and $50$~nm (c), measured at $T=80$~K. Symbols are data, curves are fits as described in the text. Insets show longitudinal MOKE hysteresis loops acquired simultaneously with the ac susceptibility. Arrows show the directions of the field sweep.}
\end{figure}

Figure~\ref{fig:suscept_80K} shows the results for $T=80$~K, for Py thicknesses $t=20$~nm, $35$~nm, and $50$~nm. The longitudinal MOKE hysteresis loops shown in the insets illustrate that the samples exhibit well-defined switching between two reversed orientations of $M$. The coercive fields $H_{C1}$, $H_{C2}$ determined from the dc MOKE hysteresis loops coincide with the peaks in the corresponding branches of ac susceptibility curves, consistent with the calculations in Fig.~\ref{fig:meas_schematic}(c). All three samples exhibit directional asymmetry of dc MOKE hysteresis loops characterized by the effective EB fields $H_E=-(H_{C1}+H_{C2})/2$ of $185$~Oe, $80$~Oe, and $62$~Oe for $t=20$~nm, $35$~nm, and $50$~nm, respectively. The EB field scales approximately inversely with $t$, consistent with averaging of the effects of F/AF interface over the Py film thickness.

Magneto-optic imaging discussed below shows that magnetization reversal occurs via propagation of domain walls (DWs) separating two oppositely magnetized states, resulting in perfectly square hysteresis loops obtained by averaging the magneto-optical images of $65~\mu m\times 65~\mu m$ sample regions [see Fig.~\ref{fig:imaging}]. Small deviations from the square shape of the hysteresis loops in Fig.~\ref{fig:suscept_80K} are thus associated with the local variations of DW nucleation and propagation thresholds in different sample regions, resulting in multidomain intermediate states of Py. Analysis of the field dependence of susceptibility described below was performed using the decreasing-field-magnitude branches of the hysteresis curves, corresponding to the quasi-uniform magnetization states.

The susceptibility data for $t=20$~nm [Fig.~\ref{fig:suscept_80K}(a)] are consistent with the presence of unidirectional anisotropy. Indeed, the asymmetry of susceptibility can be approximately described as a shift in the negative-field direction, as expected due to a positive effective EB field $H_{ud}$. The data for $t=35$~nm and $t=50$~nm [Figs.~\ref{fig:suscept_80K}(b),(c)] also exhibit a noticeable asymmetry. However, this asymmetry is predominantly associated with the hysteretic magnetization reversal. As deiscussed above, the hysteresis is likely associated with inhomogeneous states involved in the reversal process, whose response to ac field is dependent on the spatial magnetization distribution. To avoid this complication, we focus only on the dependence of susceptibility on $H_{dc}$ for the quasi-uniform magnetization states, obtained when the field magnitude is decreased from large saturating values. For these branches, the susceptibility is almost symmetric with respect to the field direction. Thus, we come to a surprising conclusion: {\it the effective unidirectional anisotropy field essentially vanishes, despite a significant asymmetry of the hysteresis loops}.

We note that for $t=20$~nm [Fig.~\ref{fig:suscept_80K}(a)], the peak of transverse susceptibility in the upward field sweep is significantly smaller than in the downward sweep. For $t=35$~nm and $50$~nm [Figs.~\ref{fig:suscept_80K}(b),(c)], the asymmetry is even more dramatic, as the peak in the upward sweep is replaced by a non-hysteretic crossover to the downward-sweep branch. We will discuss our interpretation of these behaviors in the next section.

\begin{figure}
	\includegraphics[width=0.7\columnwidth]{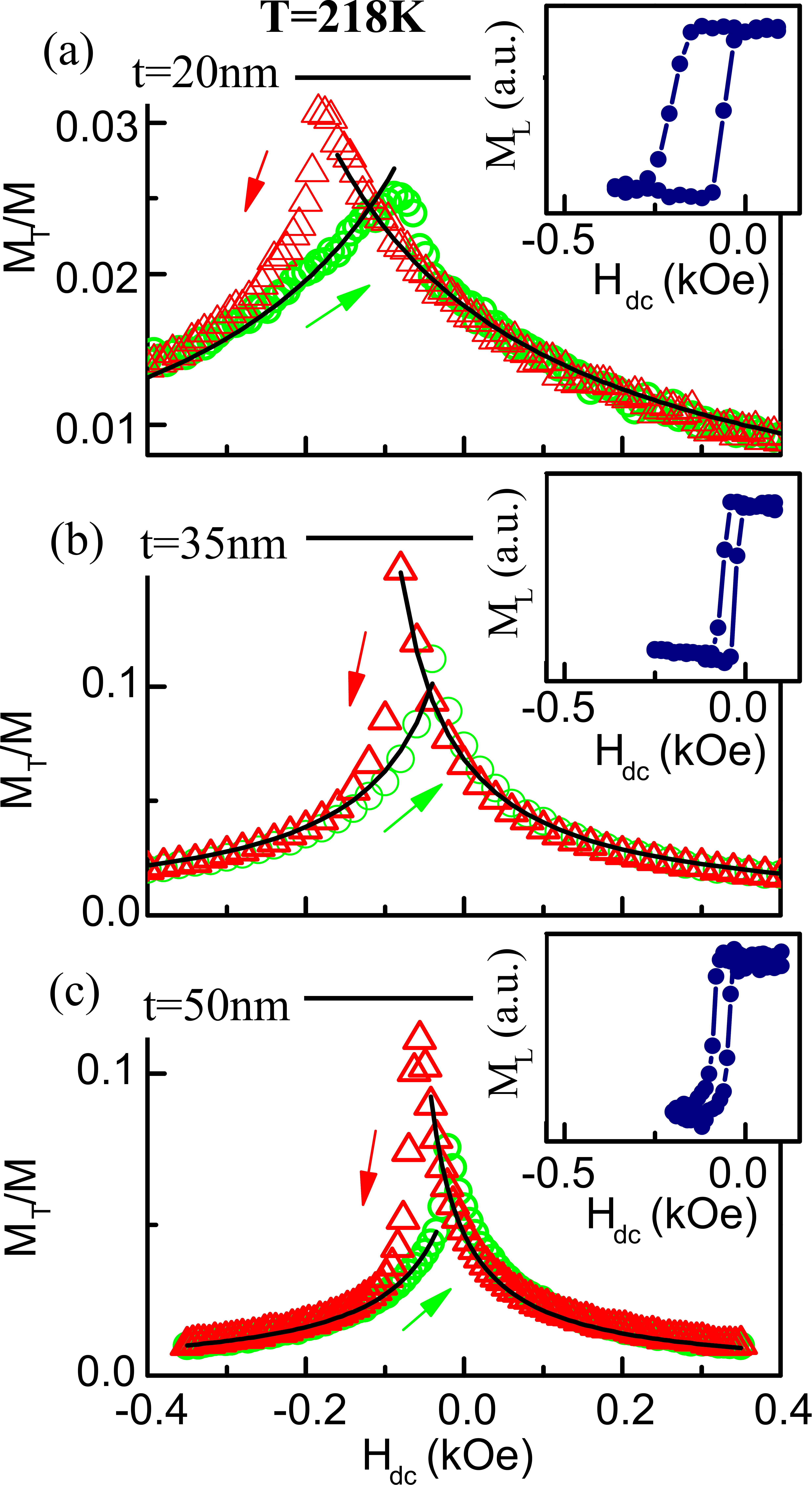}
	\caption{\label{fig:suscept_218K}(Color online) Same as Fig.~\ref{fig:suscept_80K}, at $T=218$~K.}
\end{figure}

We quantified the effective exchange fields by fitting the decreasing-field-magnitude branches of susceptibility curves with Eq.~(\ref{eq:M_T}). This equation provided excellent fitting for all the data, except for the negative-field branch for $t=20$~nm  [Fig.~\ref{fig:suscept_80K}(a)]. In the latter case, the fitting curve underestimated the susceptibility at large negative fields, and overestimated it at small fields. This discrepancy is explained by a significant inhomogeneity of the magnetization state due to the torques exerted on $\mathbf{M}$ by AF, as discussed in the next section. The values of effective unidirectional anisotropy field $H_{ud}$ obtained from fitting are $110$, $0$, and $20$~Oe for $t=20$, $35$, and $50$~nm, respectively. The vanishing of unidirectional anisotropy in thick Py is consistent with the qualitative analysis above.

For the uniaxial anisotropy field, we obtain $H_{ua}=670$, $230$, and $170$~Oe for $t=20$, $35$, and $50$~nm, respectively. Thus, in contrast to the unidirectional anisotropy, the uniaxial anisotropy does not abruptly vanish with increasing Py thickness. Its gradual decrease with increasing $t$ is consistent with the diminishing effects of AF on thicker Py films.

We have performed measurements at other values of temperature $T$, to ascertain that the observed crossover of magnetic anisotropy is a robust characteristic of the studied structures. Figure~\ref{fig:suscept_218K} shows the results for $T=218$~K, which is the base temperature of the magneto-optical imaging discussed below. This temperature is much closer to the N\`eel temperature $T_N=290$~K of CoO, and consequently the EB effects are significantly smaller than at $80$~K. Nevertheless, a shift of the longitudinal hysteresis loops in the negative-field direction is apparent. Fitting the susceptibility curves with Eq.~(\ref{eq:M_T}) yields  $H_{ud}=120$~Oe for $t=20$~nm, and $20$~Oe for $t=50$~nm. We note that the unidirectional anisotropy is almost unchanged from $T=80$~K, even though the EB is significantly smaller. In the rest of this paper, we present complementary measurements, analysis and simulations that elucidate the microscopic mechanisms of these behaviors.

\subsection{Magneto-optic imaging of magnetization reversal} 

\begin{figure}
	\includegraphics[width=\columnwidth]{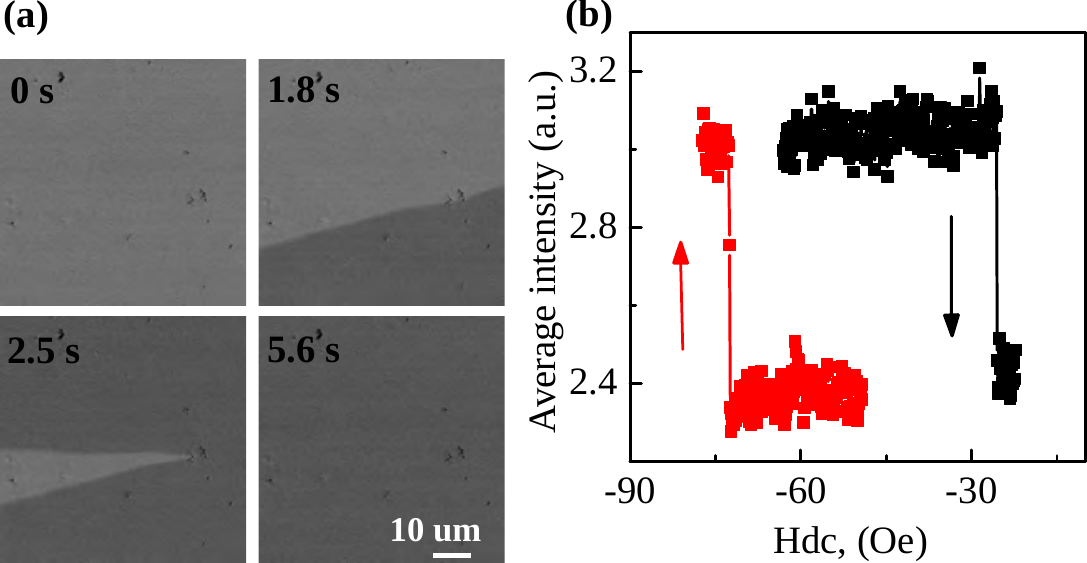}	
	\caption{\label{fig:imaging} (Color online) (a) Representative MOKE snapshots of a $65~\mu m\times 65~\mu m$ area for the Py(35)CoO(8) sample capturing its magnetization reversal, at the labeled instants of time. The images were acquired at $T=218$~K, with the horizontal dc field slowly swept at a rate of $0.14$~Oe/s from the initial value of $-23.7$~Oe. (c) Partial MOKE hysteresis loops obtained by averaging the intensities over the images such as those in (a).}
\end{figure}

To ascertain that the surprising effects observed in the ac susceptibility measurements are not associated with some complex disordered magnetization states, we performed real-time magneto-optical imaging of the magnetization reversal process with a slowly swept field. Figure~\ref{fig:imaging} shows results representative of all the studied samples, for both directions of magnetization reversal. We have repeated measurements over different sample regions, to verify that our observations are representative. For both field sweep directions, the reversal typically proceeded via a sweep of a nearly straight DW. In the less-typical instance shown in Fig.~\ref{fig:imaging}(a), an upward-propagating DW was stopped by a defect in the middle-right part of the image, and reversal proceeded via a downward-propagating DW. Nevertheless, the entire reversal was completed within $1$~Oe range of $H_{dc}$.

No significant DW roughness or distortions during propagation were observed, consistent with weak magnetization pinning by defects. The magnetization in the regions separated by the DW is uniformly saturated in the opposite directions, as is also evident from the square partial hysteresis loop obtained by averaging the MOKE intensity over the image area [Fig.~\ref{fig:suscept_218K}(c)]. Based on these results, we conclude that the studied systems remain quasi-uniformly magnetized except for reversal, validating the presented analysis of anisotropy based on the transverse  susceptibility measurements.

\subsection{Magnetoelectronic characterization of magnetization state} 

\begin{figure}
	\includegraphics[width=0.9\columnwidth]{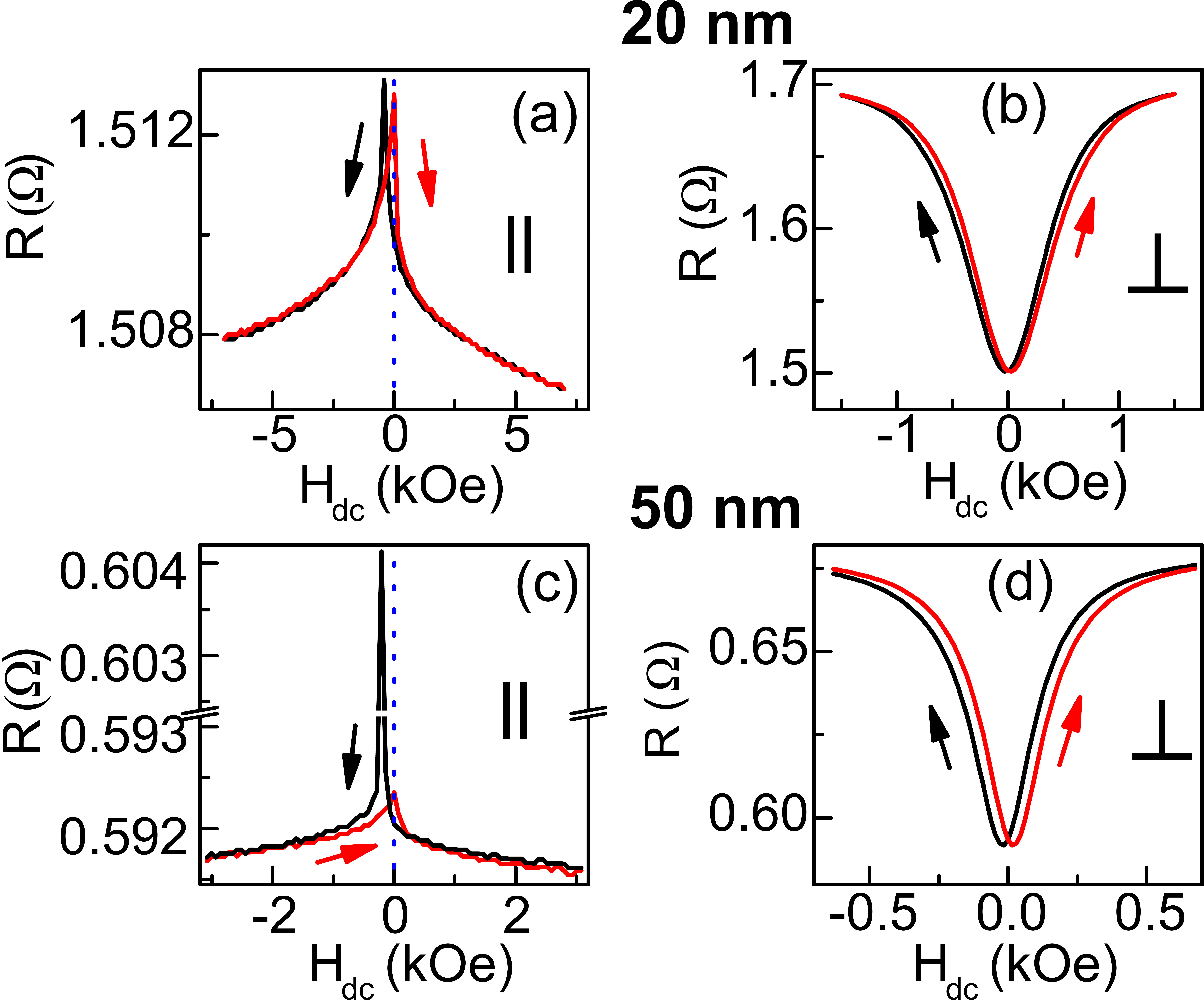}
	\caption{\label{fig:AMR} Magnetoelectronic hysteresis curves for the Py thickness $t=20$~nm (a,b) and $t=50$~nm (c,d) acquired with the field direction parallel (a,c) and perpendicular (b,d) to the cooling field, at $T=80$~K. The current flows perpendicular to the cooling field. The dotted vertical lines in (a), (c) show $H_{dc}=0$.}
\end{figure}

Magnetoelectronic measurements utilizing AMR independently confirmed the surprising dependence of anisotropy on Py thickness observed in transverse susceptibility measurements. Figure~\ref{fig:AMR} shows representative magnetoelectronic hysteresis loops for $t=20$~nm and $50$~nm acquired at $T=80$~K, the base temperature of our transverse susceptibility measurements.   

In our measurements, the current direction was orthogonal to the cooling field $H_{cool}$. Thus, in measurements with $\mathbf{H}_{dc}\parallel\mathbf{H}_{cool}$ shown in panels (a) and (c), the resistance $R$ was minimized when the magnetization was saturated in the direction of the field. Deviations from  saturation are detected as a resistance increase, which is approximately proportional to $\left<\theta^2\right>$, the average of the square of the angle between $\mathbf{M}$ and $\mathbf{H}_{dc}$. Conversely, for $\mathbf{H}_{dc}\perp \mathbf{H}_{cool}$ [panels (b) and (d)], $R$ decreases when the magnetization deviates from the direction of the field.

To facilitate direct comparison between the data for different Py thicknesses, we used $2.5$ times larger horizontal (field) scales in panels (a),(b) than in panels (c),(d), respectively. This accounted for the $2.5$ times smaller Py thickness in these panels, resulting in smaller effects of $H_{dc}$ relative to those of the effective exchange fields at F/AF interface. Similarly, we adjusted the vertical (resistance) scales to account for the different values of magnetoresistance for the two samples. Specifically, the AMR of the sample with $t=20$~nm is $0.2$~$\Omega$, while the AMR of the sample with $t=50$~nm is $0.085$~$\Omega$. Accordingly, the overall vertical scales in panels (a),(b) are $2.3$ times larger than in panels (c) and (d), correspondingly.

The appropriateness of such scaling is evident from the data for $\mathbf{H}_{dc}\perp\mathbf{H}_{cool}$, which are nearly identical for two different thicknesses. In this case, as the magnitude of $\mathbf{H}_{dc}\perp\mathbf{H}_{cool}$ is decreased, $R$ gradually and almost non-hysteretically decreases, at $H_{dc}=0$ almost reaching its minimal value achieved at large $\mathbf{H}_{dc}\parallel\mathbf{H}_{cool}$ [compare to panels (a), (c)]. Thus, at small $H_{dc}$ the magnetization forms a quasi-uniform configuration orthogonal to $H_{dc}$, consistent with the dominant effects of $H_{ua}$ and/or $H_{ud}$. 

The results for$\mathbf{H}_{dc}\parallel\mathbf{H}_{cool}$, are noticeably different for two thicknesses. For $t=20$~nm, the heights of the two peaks associated with reversal are similar, and the overall asymmetry with respect to the field direction exhibits an overall shift consistent with the presence of $H_{ud}$. In contrast, for $t=50$~nm, the peak in the downward field sweep is significantly taller than for the upward sweep. At larger $H_{dc}$, the variations of $R$ and the asymmetry between the two directions of $H_{dc}$ are much smaller than for $t=20$~nm. These results are consistent with our analysis of ac susceptibility.

\section{Analysis and Simulations}

We first summarize the key observations that do not fit in the common picture of EB as the result of an effective exchange field produced by ''frozen" AF spins at F/AF interface. We then introduce a microscopic picture that captures our observations, and present micromagnetic simulations supporting its viability.

\subsection{Qualitative analysis}

Our main new observation is that for sufficiently large Py thickness $t$, the unidirectional anisotropy vanishes, despite a sizable asymmetry of the magnetic hysteresis loop. For instance, for $t=50$~nm at $T=80$~K, magnetization reversal occurs at coercive fields $H_{C1}=-146$~Oe, $H_{C2}=22$~Oe [see inset in Fig.~\ref{fig:suscept_80K}(c)], yielding the conventionally defined EB field $H_E=-(H_{C1}+H_{C2})/2=62$~Oe, and the coercivity $H_C=(H_{C2}-H_{C1})/2=84$~Oe. These values are traditionally associated with the unidirectional and uniaxial anisotropies, respectively. However, quantitative characterization of anisotropy by transverse ac susceptibility yields effective unidirectional anisotropy field $H_{ud}=20$~Oe, more than $3$ times smaller than $H_E$, while the effective uniaxial field is $H_{ua}=170$~Oe, twice as large as the coercivity. These effects are confirmed by the analysis of the magnetoelectronic hysteresis curves, whose asymmetry for large Py thicknesses cannot be described as a shift induced by an effective field.

For $t=20$~nm, the susceptibility and the AMR data can be approximately described as a shift, but the field dependence of susceptibility significantly deviates from the fitting curve [Fig.~\ref{fig:suscept_80K}(a)]. This result cannot be explained by the spatial variations of the magnitude of effective exchange fields with the same orientation. Indeed, for some distribution $H_{eff,i}=\left<H_{eff}\right>+\Delta H_i$, where $\Delta H_i$ is the local deviation of effective field from the average $\left<H_{eff}\right>$, Taylor expansion of Eq.~(\ref{eq:M_T}) with respect to $\Delta H_i/\left<H_{eff}\right>$ yields $\left<M_T/M\right>\approx H_{ac}/\left<H_{eff}\right>+\left<\Delta H_i^2\right>/\left<H_{eff}\right>^2$, which describes enhancement of susceptibility at small fields, opposite to the observed trend. These results provide evidence for deviations of the magnetization state from saturation that become enhanced at small $H_{dc}$, resulting in reduced susceptibility. However, magneto-optical imaging shows clean switching between two oppositely magnetized saturated magnetization states. Thus, such deviations must be small and/or spatially localized to nanoscale dimensions.

We now introduce a microscopic model that accounts for these seemingly contradictory observations, as well as other puzzling features of our data. We start with the idea of Koon, Schulthess and Butler, that exchange interaction at F/AF interface results in the spin-flop in AF, producing uniaxial anisotropy~\cite{PhysRevLett.78.4865,PhysRevLett.81.4516}. Indeed, both the susceptibility and AMR measurements reveal uniaxial anisotropy whose easy axis is well-aligned with the cooling field. The stability of the anisotropy axis implies that the local N\'eel vector must be ''frozen" by the field-cooling, which is likely facilitated by the random variations of exchange coupling at F/AF interface that result in complex AF energy landscape enabling multiple metastable magnetic configurations, among them the spin-flop configurations favored by the field-cooling.

By itself, this interpretation does not explain EB, because the spin-flop state is bistable~\cite{PhysRevLett.81.4516}. To explain the existence of EB in the absence of directional anisotropy, we invoke the argument of Malozemoff, that interface imperfections frustrate exchange interaction across the A/AF interface~\cite{PhysRevB.35.3679}. This implies that upon field-cooling, the direction of AF spin-flop locally randomly varies around the average defined by the field-cooling. 

Consider now the process of the Py magnetization reversal via domain wall (DW) sweep [see Fig.~\ref{fig:imaging}]. For magnetic films with in-plane anisotropy, the DW is of the N\'eel type, with the magnetization twisted in the DW in-plane. The AF spin tilting reversal driven by the propagating DW imprints the twist direction into an asymmetry of the interfacial AF spin distribution with respect to the field direction. The effect of this asymmetric distribution on the reversed magnetization $M$ of Py is a torque rotating it in the opposite direction. Spatial fluctuations of F/AF exchange coupling result in local ''hotspots" where this torque is large, providing local nucleation centers for the reversal of magnetization back into the field-cooled direction, which is expected to occur close to $H_{dc}=0$, consistent with the data [see Figs.~\ref{fig:suscept_80K}, \ref{fig:suscept_218K}].

The proposed mechanism locally acts as an effective AF exchange spring that ``winds" during initial reversal, and whose unwinding facilitates the reversal of $M$ back into the cooling-field direction, reminiscent of the mechanism proposed by Mauri~\cite{Mauri1987}. However, if an actual exchange spring were created in CoO, its wound-state exchange energy $E_{ex}$ per unit area of the film would determine the effective unidirectional field $H_{ud}=E_{ex}/2\mu_0Mt$. In contrast, the proposed mechanism does not require a difference in magnetic energy between the two opposite magnetization states, consistent with the lack of unidirectional anisotropy in our measurements for thick Py films. Central to our interpretation of the origin of EB in the studied system is the asymmetry of the magnetization reversal process. Such asymmetry has been extensively reported for a variety of F/AF systems, suggesting its general significance for EB~\cite{mccord2003kerr,PhysRevB.73.184428,jimenez2009emergence}.

The proposed mechanism explains the observed dependence of anisotropy on the Py thickness $t$, as follows. For $t=20$~nm, the  torques exerted on $\mathbf{M}$ by AF result in its more significant twisting than in thicker Py, where the effects of spatially varying F/AF exchange become efficiently averaged through the F thickness. In turn, this implies that the torques exerted by thinner F on AF are smaller, resulting in only partial reversal of spin-flop state in AF. Thus, for thin Py, some of the AF spins remain ''frozen", in agreement with the conventional picture of EB.

The proposed mechanism also explains why the susceptibility exhibits large peaks only for the decreasing-field branches. For increasing field, the reversal of $\mathbf{M}$ facilitated by the nucleation at ''hotspots" occurs once the energy of the reversed state becomes lower, i.e., at the crossing of the $H_{dc}<0$ and $H_{dc}>$ branches of susceptibility curves [see Eq.~(\ref{eq:M_T})]. For negligible $H_{ud}$, this is expected to occur at $H_{dc}=0$, in excellent agreement with data [Fig.~\ref{fig:suscept_80K}].

\subsection{Micromagnetic simulations}

We now present micromagnetic simulations that support the proposed mechanism. The simulations were performed with Mumax3 software~\cite{Vansteenkiste2014, Lel2014, Exl2014}, using a mesh of micromagnetic cells with dimensions $2.5$~nm$\times 2.5$~nm$\times3$~nm, where the third dimension refers to the out-of-plane direction. The simulated area was $320$~nm$\times 320$~nm, with the Py layer thickness varied between $24$~nm and $48$~nm. The magnetic parameters of Py were: saturation magnetization  $M_{sat}=8\times10^5$~A/m, Gilbert damping constant $0.02$, and exchange stiffness $J_F=13\times10^{-12}$~J/m. 

\begin{figure}
	\includegraphics[width=\columnwidth]{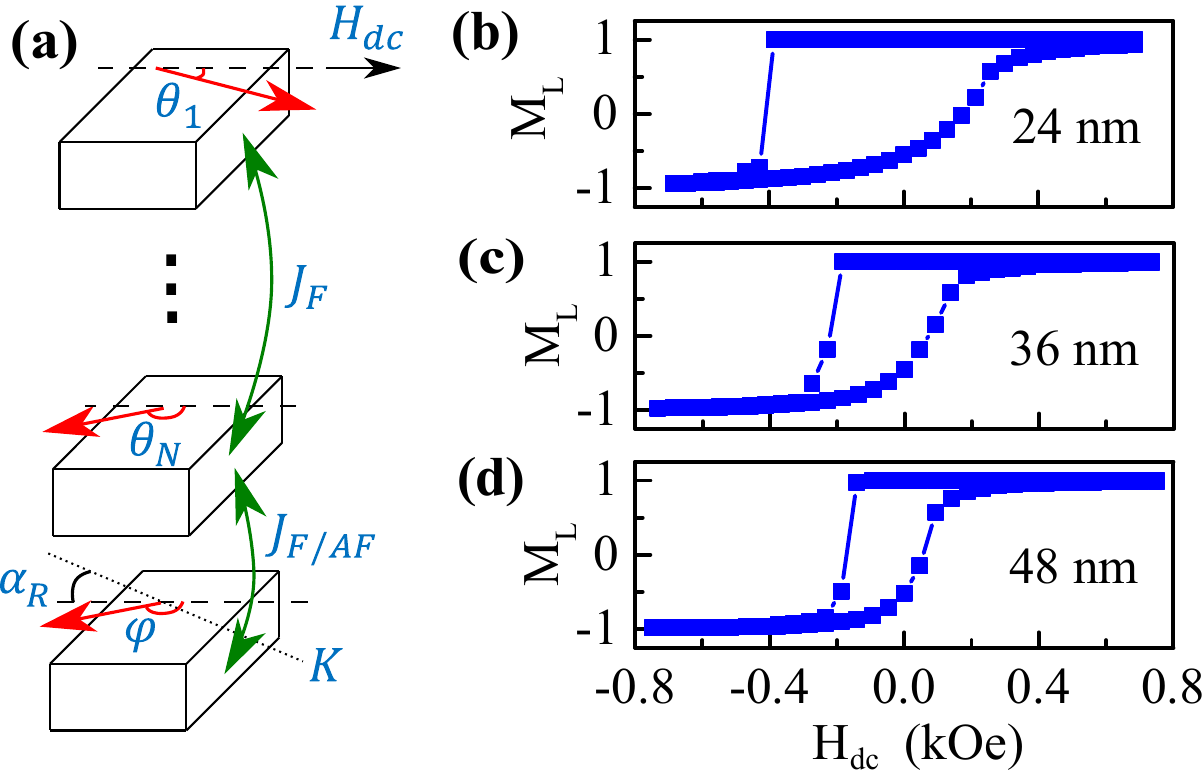}	
	\caption{\label{fig:micromagnetic} (Color online) (a) Schematic of the vertical stack of micromagnetic cells used in simulations. Straight arrows show the magnetic moments, curved arrows show the interactions. $\theta_i, i=1..N$ is angle between $H_{dc}$ and the magnetic moment of cell $i$, $\phi$ is the angle between $H_{dc}$ and magnetic moment of the cell representing AF, $J_F$ and $J_{F/AF}$ are exchange interaction constants within F and at F/AF interface, respectively, $K$ is the uniaxal anisotropy of AF layer, $\alpha_R$ - random angle between easy anisotropy axis and $H_{dc}$. (b-d) Simulated hysteresis loops for the F thickness $t=24$~nm, $36$~nm, and $48$~nm, as labeled.}
\end{figure}

The bistable spin-flop state of AF was modeled by a single layer adjacent to F, with randomly oriented uniaxial anisotropy of fixed magnitude $K = 550$~kJ/m$^{3}$, which provided a good matching with the experimental hysteresis loops. The exchange constant $J_{F/AF}$ for interaction between this layer and F was the same as within F, but the exchange between the cells representing AF was negligible. A similar approximation is utilized in the ''granular" models of EB~\cite{Fulcomer1972,PhysRevB.59.3722,OGRADY2010883}. The parameters and the interactions through the vertical stack of micormagnetic cells are schematically shown in  Fig.~\ref{fig:micromagnetic}(a). 

At the start of simulations, the magnetic moments of all the cells, including F and AF, were initialized in a uniform state with $\varphi=\theta_i=0$ [see Fig.~\ref{fig:micromagnetic} for nomenclature] that defined the ''field-cooling direction", and the system was then allowed to relax. The field $H_{dc}$ was applied at a small positive angle relative to this direction, to simulate the directional asymmetry induced by the reversal via domain wall sweep. To determine the stationary state for a given $H_{dc}$, the system was allowed to relax until its dynamics became negligible.

Panels (b-d) in Fig.~\ref{fig:micromagnetic}(b-d) show the simulated hysteresis loops for Py thicknesses $t=24$~nm, $36$~nm, and $48$~nm, close to the values utilized in measurements. All three loops exhibit directional asymmetry and enhanced coercivity which decrease with increasing $t$, consistent with the experimental observations. Furthermore, the coercive fields are in close agreement with those measured for similar values of $t$. However, the gradual reversal into the ''field-cooling" direction, and the abrupt reversal against it are opposite to the experimental trends. These differences are likely associated with the neglected effects of long-range AF correlations on the magnetic energy landscape, as well as thermal fluctuations.

\begin{figure}
	\includegraphics[width=\columnwidth]{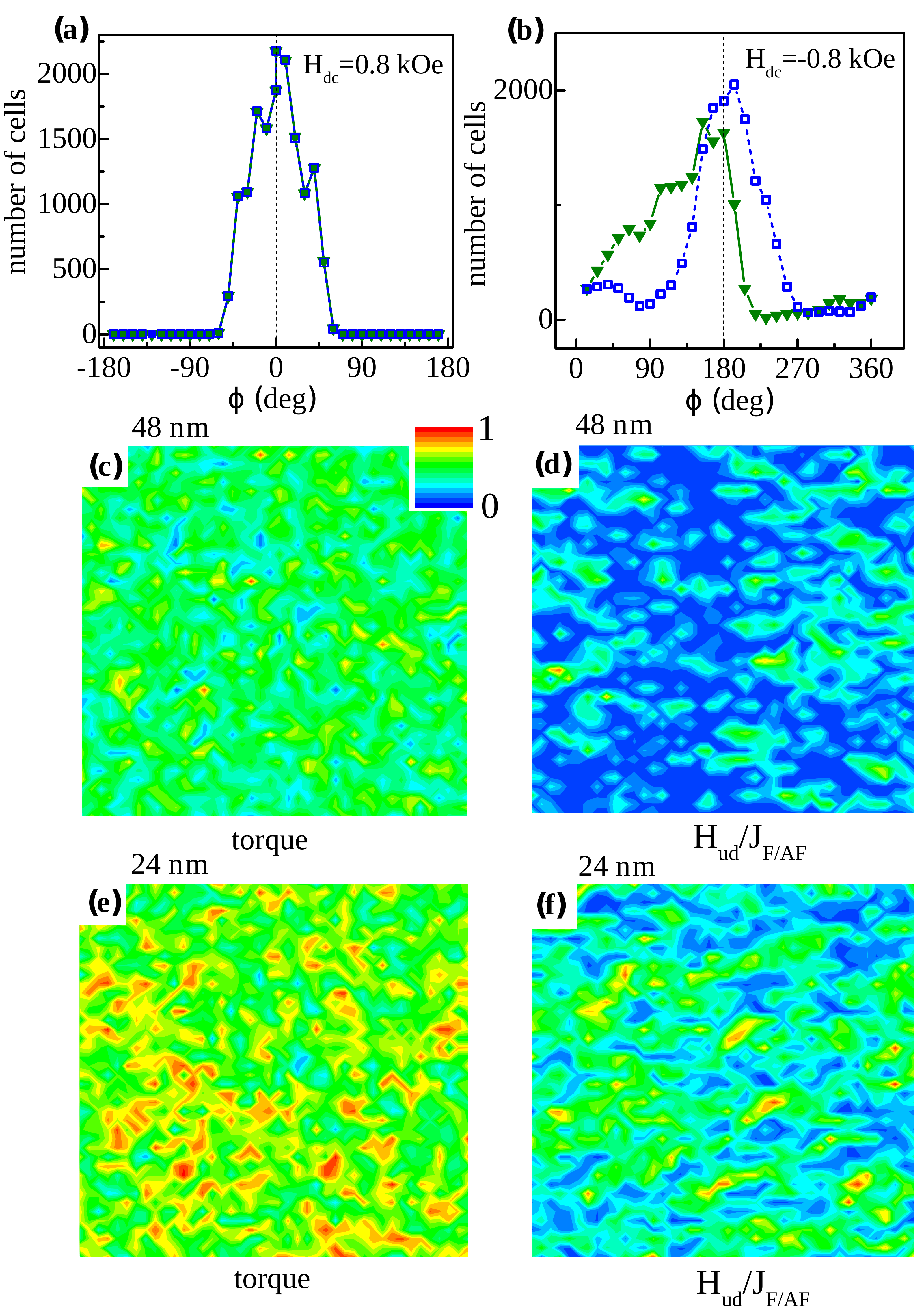}	
	\caption{\label{fig:micromagnetic_distribution} (Color online) (a) Simulated distribution of the AF magnetization angle $\varphi$ in the initial state at $H_{dc}=0.8$~kOe, for $t=24$~nm (symbols) and $t=48$~nm (curve). (b) Distribution of $\varphi$ in the reversed state at $H_{dc}=-0.8$~kOe, for $t=24$~nm (solid symbols), and $t=48$~nm (open symbols). (c) Map of torques exerted by AF on saturated $M$ in the reversed state, calculated as $|sin\varphi|$, for $t=48$~nm. (d) Map of AF-induced unidirectional anisotropy calculated as ($cos\varphi^{(i)}+cos\varphi^{(f)}$)/2, for $t=48$~nm. (e),(f) Same as (c),(d), for $t=24$~nm. The maps (c)-(f) are $320$~nm$\times 320$~nm.}
\end{figure}

We now analyze the microscopic mechanisms underlying these switching behaviors, focusing on the distribution of angle $\varphi$ between the AF spins and $\mathbf{H}_{dc}$, for thicknesses $t=24$ and $t=48$~nm. The initial distribution is centered around $\varphi=0$ and is the same for both thicknesses [(Fig.~\ref{fig:micromagnetic_distribution}(a)], as determined by the initial simulation conditions and the distribution of uniaxial anisotropy. 

In the reversed state, the distribution for $t=48$~nm is dominated by the peak around $\varphi=180^\circ$, i.e., nearly all of the AF spins become reversed and the directional anisotropy associated with the ''frozen" AF spins is negligible  [open symbols in Fig.~\ref{fig:micromagnetic_distribution}(b)]. In addition, there is a bump at small $\varphi$, but no such feature near $\varphi=360^\circ$. This directional asymmetry is associated with the rotation sense of $\mathbf{M}$ during reversal, defined by a small tilt of $H_{dc}$ relative to $H_{cool}$. A spatial map of $|sin\varphi|$, which determines the torque exerted by the AF spins on the saturated $M$, exhibits ''hotspots" that serve as nucleation centers for reversal into the field-cooled state. 

The unidirectional anisotropy of F due to the unflipped AF spins is given by ($cos\varphi^{(i)}+cos\varphi^{(f)}$)/2, where $\varphi^{(i)}$ ($\varphi^{(f)}$) is the AF spin angle in the initial (reversed) state. The corresponding map also exhibits ''hotspots",  Fig.~\ref{fig:micromagnetic_distribution}(d). However, the characteristic magnitudes are small, consistent with the experiment.

In contrast to $t=48$~nm, for $t=24$~nm the AF spin distribution forms a broad peak between $\varphi=0$ and about $200^\circ$, strongly breaking the rotational symmetry [open symbols in Fig.~\ref{fig:micromagnetic_distribution}(b)]. Just like for $t=48$~nm, the spatial distribution of torques forms ''hotspots" that serve as reversal nucleation centers [Fig.~\ref{fig:micromagnetic_distribution}(e)]. However, in this case $\varphi$ is broadly distributed over a large range of angles instead of being concentrated around $0$ and $180^\circ$. As a consequence, both the torques and the unidirectional anisotropy are substantially larger [Fig.~\ref{fig:micromagnetic_distribution}(e),(f)].

The large spatially inhomogeneous torques exerted by this AF spin distribution on $M$ result in significant deviations of the latter from the direction of $H_{dc}$. These results are consistent with our experimental data. They also explain a surprising  recent observation that the effective exchange field in CoO/Py bilayers with thin Py is spatially uncorrelated only for $\mathbf{M}\parallel \mathbf{H}_{cool}$, but is correlated on the scale of the magnetic exchange length for the opposite direction of $\mathbf{M}$~\cite{PhysRevB.101.144427}. Indeed, the distribution in Fig.~\ref{fig:micromagnetic_distribution}(b) for $t=24$~nm is strongly asymmetric, resulting in spatially correlated torques on the Py magnetization.

\section{Summary and outlook}

We have presented several complementary measurements and simulations, which revealed that the magnetization states and reversal mechanisms in a model exchange bias system, a Py/CoO bilayer, are likely dominated by the random distribution of spin-flop in CoO. Magnetization reversal becomes imprinted on this distribution, resulting in ''hotspots" of asymmetric AF spin distribution that exert local torques on the Py magnetization in exchange spring-like fashion. These ''hotspots" serve as domain wall nucleation centers, facilitating magnetization reversal into the initial field-cooled state. A striking consequence of this mechanism is the negligible unidirectional anisotropy for sufficiently thick Py films, despite a significant directional asymmetry of the magnetic hysteresis loop.

Our findings can be verified by a variety of modern techniques. For instance, the absence of unidirecitonal anisotropy can be verified as a symmetric with respect to the external field direction ferromagnetic resonance frequency. Meanwhile, the linewidth is expected to be larger for the field direction opposite to the cooling field. The ''hotspots" of local torques can be detected by nanometer-scale imaging using  magnetic force microscopy or magnetic circular dichroism, which should reveal localized twists of magnetization in the direction determined by prior reversal.

Our microscopic picture is substantially different from the known to us prior work. Nevertheless, in addition to AF spin-flop and random field effects, our picture naturally incorporates elements of "granular" and AF exchange-spring models of exchange bias. These ingredients revealed by our study of CoO/Py are likely present, in varied proportions, in other exchange bias systems. We believe that our results can provide an effective template for their quantitative characterization, as well as for developing new devices taking advantage of exchange bias. For instance, our work has shown that exchange bias reversal for switchable memory applications can avoid having to overcome the directional anisotropy. Therefore, efficient reversal can be achieved simply by localizing the mechanism driving the reversal, such as the electric field~\cite{He2010} or strain~\cite{Wu2015}, to a small ''hotspot" region of the device. 

Our findings are encouraging for the development of F/AF-based magnetic memristors~\cite{PhysRevB.78.113309,PhysRevApplied.12.044029}. It was recently suggested that an ideal memristor can be implemented based on F/AF bilayers by utilizing viscous dynamics of N\'eel order close to the AB blocking temperature $T_{B}$~\cite{doi:10.1063/5.0018411}. Similarly, a proof-of-principle implementation of F/AF bilayer-based magnetic memristor required operation close to $T_{B}$~\cite{PhysRevApplied.12.044029}, severely limiting thermal compliance of such devices. Our results indicate that these requirements can be eliminated, by taking advantage of the bistable spin-flop state of AF as an atomic-scale multilevel memory mechanism. As is evident from  Fig.~\ref{fig:micromagnetic_distribution}, the distribution of AF spin-flop states reflects the history of the F magnetization dynamics, making this system well-suited for memristive applications.

The dynamics of AF spin-flop reversal in F/AF heterostructures is likely governed by the high dynamical frequencies of AF. The energy scale involved in this dynamics is of the order of exchange energy, i.e., about $100$~mV per pair of interfacial spins. Thus, fast multistable devices are achievable in the deep nanoscale regime. Efficient implementations of such devices will likely rely on the ability to tailor the magnetic properties and the interaction strengths to enable efficient and fast driving of AF spin-flop reversal during memristor writing, while retaining long-term stability in the absence of driving.

This work was supported by the NSF award Nos. ECCS-2005786 and ECCS-1833581.

\bibliography{CoOPy-EB}

\begin{thebibliography}{53}%
\makeatletter
\providecommand \@ifxundefined [1]{%
 \@ifx{#1\undefined}
}%
\providecommand \@ifnum [1]{%
 \ifnum #1\expandafter \@firstoftwo
 \else \expandafter \@secondoftwo
 \fi
}%
\providecommand \@ifx [1]{%
 \ifx #1\expandafter \@firstoftwo
 \else \expandafter \@secondoftwo
 \fi
}%
\providecommand \natexlab [1]{#1}%
\providecommand \enquote  [1]{``#1''}%
\providecommand \bibnamefont  [1]{#1}%
\providecommand \bibfnamefont [1]{#1}%
\providecommand \citenamefont [1]{#1}%
\providecommand \href@noop [0]{\@secondoftwo}%
\providecommand \href [0]{\begingroup \@sanitize@url \@href}%
\providecommand \@href[1]{\@@startlink{#1}\@@href}%
\providecommand \@@href[1]{\endgroup#1\@@endlink}%
\providecommand \@sanitize@url [0]{\catcode `\\12\catcode `\$12\catcode
  `\&12\catcode `\#12\catcode `\^12\catcode `\_12\catcode `\%12\relax}%
\providecommand \@@startlink[1]{}%
\providecommand \@@endlink[0]{}%
\providecommand \url  [0]{\begingroup\@sanitize@url \@url }%
\providecommand \@url [1]{\endgroup\@href {#1}{\urlprefix }}%
\providecommand \urlprefix  [0]{URL }%
\providecommand \Eprint [0]{\href }%
\providecommand \doibase [0]{http://dx.doi.org/}%
\providecommand \selectlanguage [0]{\@gobble}%
\providecommand \bibinfo  [0]{\@secondoftwo}%
\providecommand \bibfield  [0]{\@secondoftwo}%
\providecommand \translation [1]{[#1]}%
\providecommand \BibitemOpen [0]{}%
\providecommand \bibitemStop [0]{}%
\providecommand \bibitemNoStop [0]{.\EOS\space}%
\providecommand \EOS [0]{\spacefactor3000\relax}%
\providecommand \BibitemShut  [1]{\csname bibitem#1\endcsname}%
\let\auto@bib@innerbib\@empty
\bibitem [{\citenamefont
  {Piramanayagam}(2012)}]{piramanayagam2012developments}%
  \BibitemOpen
  \bibfield  {author} {\bibinfo {author} {\bibfnamefont {S.~N.}\ \bibnamefont
  {Piramanayagam}},\ }\href@noop {} {\emph {\bibinfo {title} {Developments in
  data storage : materials perspective}}}\ (\bibinfo  {publisher} {Wiley IEEE
  Press},\ \bibinfo {address} {Hoboken, N.J. Salem, Mass},\ \bibinfo {year}
  {2012})\BibitemShut {NoStop}%
\bibitem [{\citenamefont {{Victora}}\ and\ \citenamefont {{Xiao
  Shen}}(2005)}]{1396176}%
  \BibitemOpen
  \bibfield  {author} {\bibinfo {author} {\bibfnamefont {R.~H.}\ \bibnamefont
  {{Victora}}}\ and\ \bibinfo {author} {\bibnamefont {{Xiao Shen}}},\ }\href
  {\doibase 10.1109/TMAG.2004.838075} {\bibfield  {journal} {\bibinfo
  {journal} {IEEE Transactions on Magnetics}\ }\textbf {\bibinfo {volume}
  {41}},\ \bibinfo {pages} {537} (\bibinfo {year} {2005})}\BibitemShut
  {NoStop}%
\bibitem [{\citenamefont {Tudu}\ and\ \citenamefont
  {Tiwari}(2017)}]{TUDU2017329}%
  \BibitemOpen
  \bibfield  {author} {\bibinfo {author} {\bibfnamefont {B.}~\bibnamefont
  {Tudu}}\ and\ \bibinfo {author} {\bibfnamefont {A.}~\bibnamefont {Tiwari}},\
  }\href {\doibase https://doi.org/10.1016/j.vacuum.2017.01.031} {\bibfield
  {journal} {\bibinfo  {journal} {Vacuum}\ }\textbf {\bibinfo {volume} {146}},\
  \bibinfo {pages} {329} (\bibinfo {year} {2017})}\BibitemShut {NoStop}%
\bibitem [{\citenamefont {Timopheev}\ \emph {et~al.}(2017)\citenamefont
  {Timopheev}, \citenamefont {Teixeira}, \citenamefont {Sousa}, \citenamefont
  {Aufret}, \citenamefont {Nguyen}, \citenamefont {Buda-Prejbeanu},
  \citenamefont {Chshiev}, \citenamefont {Sobolev},\ and\ \citenamefont
  {Dieny}}]{PhysRevB.96.014412}%
  \BibitemOpen
  \bibfield  {author} {\bibinfo {author} {\bibfnamefont {A.~A.}\ \bibnamefont
  {Timopheev}}, \bibinfo {author} {\bibfnamefont {B.~M.~S.}\ \bibnamefont
  {Teixeira}}, \bibinfo {author} {\bibfnamefont {R.~C.}\ \bibnamefont {Sousa}},
  \bibinfo {author} {\bibfnamefont {S.}~\bibnamefont {Aufret}}, \bibinfo
  {author} {\bibfnamefont {T.~N.}\ \bibnamefont {Nguyen}}, \bibinfo {author}
  {\bibfnamefont {L.~D.}\ \bibnamefont {Buda-Prejbeanu}}, \bibinfo {author}
  {\bibfnamefont {M.}~\bibnamefont {Chshiev}}, \bibinfo {author} {\bibfnamefont
  {N.~A.}\ \bibnamefont {Sobolev}}, \ and\ \bibinfo {author} {\bibfnamefont
  {B.}~\bibnamefont {Dieny}},\ }\href {\doibase 10.1103/PhysRevB.96.014412}
  {\bibfield  {journal} {\bibinfo  {journal} {Phys. Rev. B}\ }\textbf {\bibinfo
  {volume} {96}},\ \bibinfo {pages} {014412} (\bibinfo {year}
  {2017})}\BibitemShut {NoStop}%
\bibitem [{\citenamefont {Nanayakkara}\ \emph {et~al.}(2016)\citenamefont
  {Nanayakkara}, \citenamefont {Vasilevskii}, \citenamefont {Eremin},
  \citenamefont {Kolentsova}, \citenamefont {Kargin}, \citenamefont {Anferov},\
  and\ \citenamefont {Kozhanov}}]{Nanayakkara2016}%
  \BibitemOpen
  \bibfield  {author} {\bibinfo {author} {\bibfnamefont {K.}~\bibnamefont
  {Nanayakkara}}, \bibinfo {author} {\bibfnamefont {I.~S.}\ \bibnamefont
  {Vasilevskii}}, \bibinfo {author} {\bibfnamefont {I.~S.}\ \bibnamefont
  {Eremin}}, \bibinfo {author} {\bibfnamefont {O.~S.}\ \bibnamefont
  {Kolentsova}}, \bibinfo {author} {\bibfnamefont {N.~I.}\ \bibnamefont
  {Kargin}}, \bibinfo {author} {\bibfnamefont {A.}~\bibnamefont {Anferov}}, \
  and\ \bibinfo {author} {\bibfnamefont {A.}~\bibnamefont {Kozhanov}},\
  }\href@noop {} {\bibfield  {journal} {\bibinfo  {journal} {Journal of Applied
  Physics}\ }\textbf {\bibinfo {volume} {119}},\ \bibinfo {pages} {233906}
  (\bibinfo {year} {2016})}\BibitemShut {NoStop}%
\bibitem [{\citenamefont {{Wang}}\ \emph {et~al.}(2009)\citenamefont {{Wang}},
  \citenamefont {{Chen}}, \citenamefont {{Xi}}, \citenamefont {{Li}},\ and\
  \citenamefont {{Dimitrov}}}]{4781542}%
  \BibitemOpen
  \bibfield  {author} {\bibinfo {author} {\bibfnamefont {X.}~\bibnamefont
  {{Wang}}}, \bibinfo {author} {\bibfnamefont {Y.}~\bibnamefont {{Chen}}},
  \bibinfo {author} {\bibfnamefont {H.}~\bibnamefont {{Xi}}}, \bibinfo {author}
  {\bibfnamefont {H.}~\bibnamefont {{Li}}}, \ and\ \bibinfo {author}
  {\bibfnamefont {D.}~\bibnamefont {{Dimitrov}}},\ }\href {\doibase
  10.1109/LED.2008.2012270} {\bibfield  {journal} {\bibinfo  {journal} {IEEE
  Electron Device Letters}\ }\textbf {\bibinfo {volume} {30}},\ \bibinfo
  {pages} {294} (\bibinfo {year} {2009})}\BibitemShut {NoStop}%
\bibitem [{\citenamefont {Pershin}\ and\ \citenamefont
  {Di~Ventra}(2008)}]{PhysRevB.78.113309}%
  \BibitemOpen
  \bibfield  {author} {\bibinfo {author} {\bibfnamefont {Y.~V.}\ \bibnamefont
  {Pershin}}\ and\ \bibinfo {author} {\bibfnamefont {M.}~\bibnamefont
  {Di~Ventra}},\ }\href {\doibase 10.1103/PhysRevB.78.113309} {\bibfield
  {journal} {\bibinfo  {journal} {Phys. Rev. B}\ }\textbf {\bibinfo {volume}
  {78}},\ \bibinfo {pages} {113309} (\bibinfo {year} {2008})}\BibitemShut
  {NoStop}%
\bibitem [{\citenamefont {Krzysteczko}\ \emph {et~al.}(2009)\citenamefont
  {Krzysteczko}, \citenamefont {Reiss},\ and\ \citenamefont
  {Thomas}}]{Krzysteczko2009}%
  \BibitemOpen
  \bibfield  {author} {\bibinfo {author} {\bibfnamefont {P.}~\bibnamefont
  {Krzysteczko}}, \bibinfo {author} {\bibfnamefont {G.}~\bibnamefont {Reiss}},
  \ and\ \bibinfo {author} {\bibfnamefont {A.}~\bibnamefont {Thomas}},\ }\href
  {\doibase 10.1063/1.3224193} {\bibfield  {journal} {\bibinfo  {journal}
  {Applied Physics Letters}\ }\textbf {\bibinfo {volume} {95}},\ \bibinfo
  {pages} {112508} (\bibinfo {year} {2009})}\BibitemShut {NoStop}%
\bibitem [{\citenamefont {Krzysteczko}\ \emph {et~al.}(2012)\citenamefont
  {Krzysteczko}, \citenamefont {M\"{u}nchenberger}, \citenamefont
  {Sch\"{a}fers}, \citenamefont {Reiss},\ and\ \citenamefont
  {Thomas}}]{Krzysteczko2012}%
  \BibitemOpen
  \bibfield  {author} {\bibinfo {author} {\bibfnamefont {P.}~\bibnamefont
  {Krzysteczko}}, \bibinfo {author} {\bibfnamefont {J.}~\bibnamefont
  {M\"{u}nchenberger}}, \bibinfo {author} {\bibfnamefont {M.}~\bibnamefont
  {Sch\"{a}fers}}, \bibinfo {author} {\bibfnamefont {G.}~\bibnamefont {Reiss}},
  \ and\ \bibinfo {author} {\bibfnamefont {A.}~\bibnamefont {Thomas}},\ }\href
  {\doibase 10.1002/adma.201103723} {\bibfield  {journal} {\bibinfo  {journal}
  {Advanced Materials}\ }\textbf {\bibinfo {volume} {24}},\ \bibinfo {pages}
  {762} (\bibinfo {year} {2012})}\BibitemShut {NoStop}%
\bibitem [{\citenamefont {Balents}(2010)}]{Balents2010}%
  \BibitemOpen
  \bibfield  {author} {\bibinfo {author} {\bibfnamefont {L.}~\bibnamefont
  {Balents}},\ }\href {\doibase 10.1038/nature08917} {\bibfield  {journal}
  {\bibinfo  {journal} {Nature}\ }\textbf {\bibinfo {volume} {464}},\ \bibinfo
  {pages} {199} (\bibinfo {year} {2010})}\BibitemShut {NoStop}%
\bibitem [{\citenamefont {Binder}\ and\ \citenamefont
  {Young}(1986)}]{RevModPhys.58.801}%
  \BibitemOpen
  \bibfield  {author} {\bibinfo {author} {\bibfnamefont {K.}~\bibnamefont
  {Binder}}\ and\ \bibinfo {author} {\bibfnamefont {A.~P.}\ \bibnamefont
  {Young}},\ }\href {\doibase 10.1103/RevModPhys.58.801} {\bibfield  {journal}
  {\bibinfo  {journal} {Rev. Mod. Phys.}\ }\textbf {\bibinfo {volume} {58}},\
  \bibinfo {pages} {801} (\bibinfo {year} {1986})}\BibitemShut {NoStop}%
\bibitem [{\citenamefont {Chen}\ \emph
  {et~al.}(2020{\natexlab{a}})\citenamefont {Chen}, \citenamefont {Ivanov},\
  and\ \citenamefont {Urazhdin}}]{doi:10.1063/5.0018411}%
  \BibitemOpen
  \bibfield  {author} {\bibinfo {author} {\bibfnamefont {G.}~\bibnamefont
  {Chen}}, \bibinfo {author} {\bibfnamefont {S.}~\bibnamefont {Ivanov}}, \ and\
  \bibinfo {author} {\bibfnamefont {S.}~\bibnamefont {Urazhdin}},\ }\href
  {\doibase 10.1063/5.0018411} {\bibfield  {journal} {\bibinfo  {journal}
  {Applied Physics Letters}\ }\textbf {\bibinfo {volume} {117}},\ \bibinfo
  {pages} {103501} (\bibinfo {year} {2020}{\natexlab{a}})}\BibitemShut
  {NoStop}%
\bibitem [{\citenamefont {Mansueto}\ \emph {et~al.}(2019)\citenamefont
  {Mansueto}, \citenamefont {Chavent}, \citenamefont {Auffret}, \citenamefont
  {Joumard}, \citenamefont {Nath}, \citenamefont {Miron}, \citenamefont
  {Ebels}, \citenamefont {Sousa}, \citenamefont {Buda-Prejbeanu}, \citenamefont
  {Prejbeanu},\ and\ \citenamefont {Dieny}}]{PhysRevApplied.12.044029}%
  \BibitemOpen
  \bibfield  {author} {\bibinfo {author} {\bibfnamefont {M.}~\bibnamefont
  {Mansueto}}, \bibinfo {author} {\bibfnamefont {A.}~\bibnamefont {Chavent}},
  \bibinfo {author} {\bibfnamefont {S.}~\bibnamefont {Auffret}}, \bibinfo
  {author} {\bibfnamefont {I.}~\bibnamefont {Joumard}}, \bibinfo {author}
  {\bibfnamefont {J.}~\bibnamefont {Nath}}, \bibinfo {author} {\bibfnamefont
  {I.}~\bibnamefont {Miron}}, \bibinfo {author} {\bibfnamefont
  {U.}~\bibnamefont {Ebels}}, \bibinfo {author} {\bibfnamefont
  {R.}~\bibnamefont {Sousa}}, \bibinfo {author} {\bibfnamefont
  {L.}~\bibnamefont {Buda-Prejbeanu}}, \bibinfo {author} {\bibfnamefont
  {I.}~\bibnamefont {Prejbeanu}}, \ and\ \bibinfo {author} {\bibfnamefont
  {B.}~\bibnamefont {Dieny}},\ }\href {\doibase
  10.1103/PhysRevApplied.12.044029} {\bibfield  {journal} {\bibinfo  {journal}
  {Phys. Rev. Applied}\ }\textbf {\bibinfo {volume} {12}},\ \bibinfo {pages}
  {044029} (\bibinfo {year} {2019})}\BibitemShut {NoStop}%
\bibitem [{\citenamefont {Malozemoff}(1987)}]{PhysRevB.35.3679}%
  \BibitemOpen
  \bibfield  {author} {\bibinfo {author} {\bibfnamefont {A.~P.}\ \bibnamefont
  {Malozemoff}},\ }\href {\doibase 10.1103/PhysRevB.35.3679} {\bibfield
  {journal} {\bibinfo  {journal} {Phys. Rev. B}\ }\textbf {\bibinfo {volume}
  {35}},\ \bibinfo {pages} {3679} (\bibinfo {year} {1987})}\BibitemShut
  {NoStop}%
\bibitem [{\citenamefont {Chen}\ \emph
  {et~al.}(2020{\natexlab{b}})\citenamefont {Chen}, \citenamefont {Collette},\
  and\ \citenamefont {Urazhdin}}]{PhysRevB.101.144427}%
  \BibitemOpen
  \bibfield  {author} {\bibinfo {author} {\bibfnamefont {G.}~\bibnamefont
  {Chen}}, \bibinfo {author} {\bibfnamefont {D.}~\bibnamefont {Collette}}, \
  and\ \bibinfo {author} {\bibfnamefont {S.}~\bibnamefont {Urazhdin}},\ }\href
  {\doibase 10.1103/PhysRevB.101.144427} {\bibfield  {journal} {\bibinfo
  {journal} {Phys. Rev. B}\ }\textbf {\bibinfo {volume} {101}},\ \bibinfo
  {pages} {144427} (\bibinfo {year} {2020}{\natexlab{b}})}\BibitemShut
  {NoStop}%
\bibitem [{\citenamefont {Urazhdin}\ and\ \citenamefont
  {Danilenko}(2015)}]{PhysRevB.92.174416}%
  \BibitemOpen
  \bibfield  {author} {\bibinfo {author} {\bibfnamefont {S.}~\bibnamefont
  {Urazhdin}}\ and\ \bibinfo {author} {\bibfnamefont {U.}~\bibnamefont
  {Danilenko}},\ }\href {\doibase 10.1103/PhysRevB.92.174416} {\bibfield
  {journal} {\bibinfo  {journal} {Phys. Rev. B}\ }\textbf {\bibinfo {volume}
  {92}},\ \bibinfo {pages} {174416} (\bibinfo {year} {2015})}\BibitemShut
  {NoStop}%
\bibitem [{\citenamefont {Ma}\ and\ \citenamefont
  {Urazhdin}(2018)}]{PhysRevB.97.054402}%
  \BibitemOpen
  \bibfield  {author} {\bibinfo {author} {\bibfnamefont {T.}~\bibnamefont
  {Ma}}\ and\ \bibinfo {author} {\bibfnamefont {S.}~\bibnamefont {Urazhdin}},\
  }\href {\doibase 10.1103/PhysRevB.97.054402} {\bibfield  {journal} {\bibinfo
  {journal} {Phys. Rev. B}\ }\textbf {\bibinfo {volume} {97}},\ \bibinfo
  {pages} {054402} (\bibinfo {year} {2018})}\BibitemShut {NoStop}%
\bibitem [{\citenamefont {Ma}\ \emph {et~al.}(2016)\citenamefont {Ma},
  \citenamefont {Cheng}, \citenamefont {Boettcher}, \citenamefont {Urazhdin},\
  and\ \citenamefont {Novozhilova}}]{PhysRevB.94.024422}%
  \BibitemOpen
  \bibfield  {author} {\bibinfo {author} {\bibfnamefont {T.}~\bibnamefont
  {Ma}}, \bibinfo {author} {\bibfnamefont {X.}~\bibnamefont {Cheng}}, \bibinfo
  {author} {\bibfnamefont {S.}~\bibnamefont {Boettcher}}, \bibinfo {author}
  {\bibfnamefont {S.}~\bibnamefont {Urazhdin}}, \ and\ \bibinfo {author}
  {\bibfnamefont {L.}~\bibnamefont {Novozhilova}},\ }\href {\doibase
  10.1103/PhysRevB.94.024422} {\bibfield  {journal} {\bibinfo  {journal} {Phys.
  Rev. B}\ }\textbf {\bibinfo {volume} {94}},\ \bibinfo {pages} {024422}
  (\bibinfo {year} {2016})}\BibitemShut {NoStop}%
\bibitem [{\citenamefont {Meiklejohn}\ and\ \citenamefont
  {Bean}(1957)}]{PhysRev.105.904}%
  \BibitemOpen
  \bibfield  {author} {\bibinfo {author} {\bibfnamefont {W.~H.}\ \bibnamefont
  {Meiklejohn}}\ and\ \bibinfo {author} {\bibfnamefont {C.~P.}\ \bibnamefont
  {Bean}},\ }\href {\doibase 10.1103/PhysRev.105.904} {\bibfield  {journal}
  {\bibinfo  {journal} {Phys. Rev.}\ }\textbf {\bibinfo {volume} {105}},\
  \bibinfo {pages} {904} (\bibinfo {year} {1957})}\BibitemShut {NoStop}%
\bibitem [{\citenamefont {Nogués}\ and\ \citenamefont
  {Schuller}(1999)}]{NOGUES1999203}%
  \BibitemOpen
  \bibfield  {author} {\bibinfo {author} {\bibfnamefont {J.}~\bibnamefont
  {Nogués}}\ and\ \bibinfo {author} {\bibfnamefont {I.~K.}\ \bibnamefont
  {Schuller}},\ }\href {\doibase https://doi.org/10.1016/S0304-8853(98)00266-2}
  {\bibfield  {journal} {\bibinfo  {journal} {Journal of Magnetism and Magnetic
  Materials}\ }\textbf {\bibinfo {volume} {192}},\ \bibinfo {pages} {203}
  (\bibinfo {year} {1999})}\BibitemShut {NoStop}%
\bibitem [{\citenamefont {Kiwi}(2001)}]{KIWI2001584}%
  \BibitemOpen
  \bibfield  {author} {\bibinfo {author} {\bibfnamefont {M.}~\bibnamefont
  {Kiwi}},\ }\href {\doibase https://doi.org/10.1016/S0304-8853(01)00421-8}
  {\bibfield  {journal} {\bibinfo  {journal} {Journal of Magnetism and Magnetic
  Materials}\ }\textbf {\bibinfo {volume} {234}},\ \bibinfo {pages} {584}
  (\bibinfo {year} {2001})}\BibitemShut {NoStop}%
\bibitem [{\citenamefont {Nogu{\'{e}}s}\ \emph {et~al.}(2005)\citenamefont
  {Nogu{\'{e}}s}, \citenamefont {Sort}, \citenamefont {Langlais}, \citenamefont
  {Skumryev}, \citenamefont {Suri{\~{n}}ach}, \citenamefont {Mu{\~{n}}oz},\
  and\ \citenamefont {Bar{\'{o}}}}]{Nogus2005}%
  \BibitemOpen
  \bibfield  {author} {\bibinfo {author} {\bibfnamefont {J.}~\bibnamefont
  {Nogu{\'{e}}s}}, \bibinfo {author} {\bibfnamefont {J.}~\bibnamefont {Sort}},
  \bibinfo {author} {\bibfnamefont {V.}~\bibnamefont {Langlais}}, \bibinfo
  {author} {\bibfnamefont {V.}~\bibnamefont {Skumryev}}, \bibinfo {author}
  {\bibfnamefont {S.}~\bibnamefont {Suri{\~{n}}ach}}, \bibinfo {author}
  {\bibfnamefont {J.}~\bibnamefont {Mu{\~{n}}oz}}, \ and\ \bibinfo {author}
  {\bibfnamefont {M.}~\bibnamefont {Bar{\'{o}}}},\ }\href {\doibase
  10.1016/j.physrep.2005.08.004} {\bibfield  {journal} {\bibinfo  {journal}
  {Physics Reports}\ }\textbf {\bibinfo {volume} {422}},\ \bibinfo {pages} {65}
  (\bibinfo {year} {2005})}\BibitemShut {NoStop}%
\bibitem [{\citenamefont {Stamps}(2000)}]{Stamps_2000}%
  \BibitemOpen
  \bibfield  {author} {\bibinfo {author} {\bibfnamefont {R.~L.}\ \bibnamefont
  {Stamps}},\ }\href {\doibase 10.1088/0022-3727/33/23/201} {\bibfield
  {journal} {\bibinfo  {journal} {Journal of Physics D: Applied Physics}\
  }\textbf {\bibinfo {volume} {33}},\ \bibinfo {pages} {R247} (\bibinfo {year}
  {2000})}\BibitemShut {NoStop}%
\bibitem [{\citenamefont {Berkowitz}\ and\ \citenamefont
  {Takano}(1999)}]{BERKOWITZ1999552}%
  \BibitemOpen
  \bibfield  {author} {\bibinfo {author} {\bibfnamefont {A.}~\bibnamefont
  {Berkowitz}}\ and\ \bibinfo {author} {\bibfnamefont {K.}~\bibnamefont
  {Takano}},\ }\href {\doibase https://doi.org/10.1016/S0304-8853(99)00453-9}
  {\bibfield  {journal} {\bibinfo  {journal} {Journal of Magnetism and Magnetic
  Materials}\ }\textbf {\bibinfo {volume} {200}},\ \bibinfo {pages} {552 }
  (\bibinfo {year} {1999})}\BibitemShut {NoStop}%
\bibitem [{\citenamefont {Giri}\ \emph {et~al.}(2011)\citenamefont {Giri},
  \citenamefont {Patra},\ and\ \citenamefont {Majumdar}}]{Giri_2011}%
  \BibitemOpen
  \bibfield  {author} {\bibinfo {author} {\bibfnamefont {S.}~\bibnamefont
  {Giri}}, \bibinfo {author} {\bibfnamefont {M.}~\bibnamefont {Patra}}, \ and\
  \bibinfo {author} {\bibfnamefont {S.}~\bibnamefont {Majumdar}},\ }\href
  {\doibase 10.1088/0953-8984/23/7/073201} {\bibfield  {journal} {\bibinfo
  {journal} {Journal of Physics: Condensed Matter}\ }\textbf {\bibinfo {volume}
  {23}},\ \bibinfo {pages} {073201} (\bibinfo {year} {2011})}\BibitemShut
  {NoStop}%
\bibitem [{\citenamefont {Mauri}\ \emph {et~al.}(1987)\citenamefont {Mauri},
  \citenamefont {Siegmann}, \citenamefont {Bagus},\ and\ \citenamefont
  {Kay}}]{Mauri1987}%
  \BibitemOpen
  \bibfield  {author} {\bibinfo {author} {\bibfnamefont {D.}~\bibnamefont
  {Mauri}}, \bibinfo {author} {\bibfnamefont {H.~C.}\ \bibnamefont {Siegmann}},
  \bibinfo {author} {\bibfnamefont {P.~S.}\ \bibnamefont {Bagus}}, \ and\
  \bibinfo {author} {\bibfnamefont {E.}~\bibnamefont {Kay}},\ }\href {\doibase
  10.1063/1.339367} {\bibfield  {journal} {\bibinfo  {journal} {Journal of
  Applied Physics}\ }\textbf {\bibinfo {volume} {62}},\ \bibinfo {pages} {3047}
  (\bibinfo {year} {1987})}\BibitemShut {NoStop}%
\bibitem [{\citenamefont {Ohldag}\ \emph {et~al.}(2003)\citenamefont {Ohldag},
  \citenamefont {Scholl}, \citenamefont {Nolting}, \citenamefont {Arenholz},
  \citenamefont {Maat}, \citenamefont {Young}, \citenamefont {Carey},\ and\
  \citenamefont {St\"ohr}}]{PhysRevLett.91.017203}%
  \BibitemOpen
  \bibfield  {author} {\bibinfo {author} {\bibfnamefont {H.}~\bibnamefont
  {Ohldag}}, \bibinfo {author} {\bibfnamefont {A.}~\bibnamefont {Scholl}},
  \bibinfo {author} {\bibfnamefont {F.}~\bibnamefont {Nolting}}, \bibinfo
  {author} {\bibfnamefont {E.}~\bibnamefont {Arenholz}}, \bibinfo {author}
  {\bibfnamefont {S.}~\bibnamefont {Maat}}, \bibinfo {author} {\bibfnamefont
  {A.~T.}\ \bibnamefont {Young}}, \bibinfo {author} {\bibfnamefont
  {M.}~\bibnamefont {Carey}}, \ and\ \bibinfo {author} {\bibfnamefont
  {J.}~\bibnamefont {St\"ohr}},\ }\href {\doibase
  10.1103/PhysRevLett.91.017203} {\bibfield  {journal} {\bibinfo  {journal}
  {Phys. Rev. Lett.}\ }\textbf {\bibinfo {volume} {91}},\ \bibinfo {pages}
  {017203} (\bibinfo {year} {2003})}\BibitemShut {NoStop}%
\bibitem [{\citenamefont {Scholl}\ \emph {et~al.}(2004)\citenamefont {Scholl},
  \citenamefont {Liberati}, \citenamefont {Arenholz}, \citenamefont {Ohldag},\
  and\ \citenamefont {St\"ohr}}]{PhysRevLett.92.247201}%
  \BibitemOpen
  \bibfield  {author} {\bibinfo {author} {\bibfnamefont {A.}~\bibnamefont
  {Scholl}}, \bibinfo {author} {\bibfnamefont {M.}~\bibnamefont {Liberati}},
  \bibinfo {author} {\bibfnamefont {E.}~\bibnamefont {Arenholz}}, \bibinfo
  {author} {\bibfnamefont {H.}~\bibnamefont {Ohldag}}, \ and\ \bibinfo {author}
  {\bibfnamefont {J.}~\bibnamefont {St\"ohr}},\ }\href {\doibase
  10.1103/PhysRevLett.92.247201} {\bibfield  {journal} {\bibinfo  {journal}
  {Phys. Rev. Lett.}\ }\textbf {\bibinfo {volume} {92}},\ \bibinfo {pages}
  {247201} (\bibinfo {year} {2004})}\BibitemShut {NoStop}%
\bibitem [{\citenamefont {Imry}\ and\ \citenamefont
  {Ma}(1975)}]{PhysRevLett.35.1399}%
  \BibitemOpen
  \bibfield  {author} {\bibinfo {author} {\bibfnamefont {Y.}~\bibnamefont
  {Imry}}\ and\ \bibinfo {author} {\bibfnamefont {S.-k.}\ \bibnamefont {Ma}},\
  }\href {\doibase 10.1103/PhysRevLett.35.1399} {\bibfield  {journal} {\bibinfo
   {journal} {Phys. Rev. Lett.}\ }\textbf {\bibinfo {volume} {35}},\ \bibinfo
  {pages} {1399} (\bibinfo {year} {1975})}\BibitemShut {NoStop}%
\bibitem [{\citenamefont {Araújo}\ \emph {et~al.}(2002)\citenamefont
  {Araújo}, \citenamefont {Machado}, \citenamefont {Rodrigues}, \citenamefont
  {Azevedo}, \citenamefont {de~Aguiar}, \citenamefont {de~Almeida},
  \citenamefont {Rezende},\ and\ \citenamefont {Egelhoff}}]{Rezende}%
  \BibitemOpen
  \bibfield  {author} {\bibinfo {author} {\bibfnamefont {A.~E. P.~d.}\
  \bibnamefont {Araújo}}, \bibinfo {author} {\bibfnamefont {F.~L.~A.}\
  \bibnamefont {Machado}}, \bibinfo {author} {\bibfnamefont {A.~R.}\
  \bibnamefont {Rodrigues}}, \bibinfo {author} {\bibfnamefont {A.}~\bibnamefont
  {Azevedo}}, \bibinfo {author} {\bibfnamefont {F.~M.}\ \bibnamefont
  {de~Aguiar}}, \bibinfo {author} {\bibfnamefont {J.~R.~L.}\ \bibnamefont
  {de~Almeida}}, \bibinfo {author} {\bibfnamefont {S.~M.}\ \bibnamefont
  {Rezende}}, \ and\ \bibinfo {author} {\bibfnamefont {W.~F.}\ \bibnamefont
  {Egelhoff}},\ }\href {\doibase 10.1063/1.1447496} {\bibfield  {journal}
  {\bibinfo  {journal} {Journal of Applied Physics}\ }\textbf {\bibinfo
  {volume} {91}},\ \bibinfo {pages} {7754} (\bibinfo {year}
  {2002})}\BibitemShut {NoStop}%
\bibitem [{\citenamefont {Jim{\'e}nez}\ \emph {et~al.}(2009)\citenamefont
  {Jim{\'e}nez}, \citenamefont {Camarero}, \citenamefont {Sort}, \citenamefont
  {Nogu{\'e}s}, \citenamefont {Mikuszeit}, \citenamefont
  {Garc{\'\i}a-Mart{\'\i}n}, \citenamefont {Hoffmann}, \citenamefont {Dieny},\
  and\ \citenamefont {Miranda}}]{jimenez2009emergence}%
  \BibitemOpen
  \bibfield  {author} {\bibinfo {author} {\bibfnamefont {E.}~\bibnamefont
  {Jim{\'e}nez}}, \bibinfo {author} {\bibfnamefont {J.}~\bibnamefont
  {Camarero}}, \bibinfo {author} {\bibfnamefont {J.}~\bibnamefont {Sort}},
  \bibinfo {author} {\bibfnamefont {J.}~\bibnamefont {Nogu{\'e}s}}, \bibinfo
  {author} {\bibfnamefont {N.}~\bibnamefont {Mikuszeit}}, \bibinfo {author}
  {\bibfnamefont {J.~M.}\ \bibnamefont {Garc{\'\i}a-Mart{\'\i}n}}, \bibinfo
  {author} {\bibfnamefont {A.}~\bibnamefont {Hoffmann}}, \bibinfo {author}
  {\bibfnamefont {B.}~\bibnamefont {Dieny}}, \ and\ \bibinfo {author}
  {\bibfnamefont {R.}~\bibnamefont {Miranda}},\ }\href@noop {} {\bibfield
  {journal} {\bibinfo  {journal} {Physical Review B}\ }\textbf {\bibinfo
  {volume} {80}},\ \bibinfo {pages} {014415} (\bibinfo {year}
  {2009})}\BibitemShut {NoStop}%
\bibitem [{\citenamefont {Malozemoff}(1988)}]{PhysRevB.37.7673}%
  \BibitemOpen
  \bibfield  {author} {\bibinfo {author} {\bibfnamefont {A.~P.}\ \bibnamefont
  {Malozemoff}},\ }\href {\doibase 10.1103/PhysRevB.37.7673} {\bibfield
  {journal} {\bibinfo  {journal} {Phys. Rev. B}\ }\textbf {\bibinfo {volume}
  {37}},\ \bibinfo {pages} {7673} (\bibinfo {year} {1988})}\BibitemShut
  {NoStop}%
\bibitem [{\citenamefont {Urazhdin}\ \emph {et~al.}(2019)\citenamefont
  {Urazhdin}, \citenamefont {Li},\ and\ \citenamefont
  {Novozhilova}}]{urazhdin2019JMMM}%
  \BibitemOpen
  \bibfield  {author} {\bibinfo {author} {\bibfnamefont {S.}~\bibnamefont
  {Urazhdin}}, \bibinfo {author} {\bibfnamefont {W.}~\bibnamefont {Li}}, \ and\
  \bibinfo {author} {\bibfnamefont {L.}~\bibnamefont {Novozhilova}},\
  }\href@noop {} {\bibfield  {journal} {\bibinfo  {journal} {Journal of
  Magnetism and Magnetic Materials}\ }\textbf {\bibinfo {volume} {476}},\
  \bibinfo {pages} {75} (\bibinfo {year} {2019})}\BibitemShut {NoStop}%
\bibitem [{\citenamefont {Koon}(1997)}]{PhysRevLett.78.4865}%
  \BibitemOpen
  \bibfield  {author} {\bibinfo {author} {\bibfnamefont {N.~C.}\ \bibnamefont
  {Koon}},\ }\href {\doibase 10.1103/PhysRevLett.78.4865} {\bibfield  {journal}
  {\bibinfo  {journal} {Phys. Rev. Lett.}\ }\textbf {\bibinfo {volume} {78}},\
  \bibinfo {pages} {4865} (\bibinfo {year} {1997})}\BibitemShut {NoStop}%
\bibitem [{\citenamefont {Schulthess}\ and\ \citenamefont
  {Butler}(1998)}]{PhysRevLett.81.4516}%
  \BibitemOpen
  \bibfield  {author} {\bibinfo {author} {\bibfnamefont {T.~C.}\ \bibnamefont
  {Schulthess}}\ and\ \bibinfo {author} {\bibfnamefont {W.~H.}\ \bibnamefont
  {Butler}},\ }\href {\doibase 10.1103/PhysRevLett.81.4516} {\bibfield
  {journal} {\bibinfo  {journal} {Phys. Rev. Lett.}\ }\textbf {\bibinfo
  {volume} {81}},\ \bibinfo {pages} {4516} (\bibinfo {year}
  {1998})}\BibitemShut {NoStop}%
\bibitem [{\citenamefont {Urazhdin}\ \emph {et~al.}(2008)\citenamefont
  {Urazhdin}, \citenamefont {Tabor},\ and\ \citenamefont
  {Lim}}]{PhysRevB.78.052403}%
  \BibitemOpen
  \bibfield  {author} {\bibinfo {author} {\bibfnamefont {S.}~\bibnamefont
  {Urazhdin}}, \bibinfo {author} {\bibfnamefont {P.}~\bibnamefont {Tabor}}, \
  and\ \bibinfo {author} {\bibfnamefont {W.-L.}\ \bibnamefont {Lim}},\ }\href
  {\doibase 10.1103/PhysRevB.78.052403} {\bibfield  {journal} {\bibinfo
  {journal} {Phys. Rev. B}\ }\textbf {\bibinfo {volume} {78}},\ \bibinfo
  {pages} {052403} (\bibinfo {year} {2008})}\BibitemShut {NoStop}%
\bibitem [{\citenamefont {Spinu}\ \emph {et~al.}(2000)\citenamefont {Spinu},
  \citenamefont {Srikanth}, \citenamefont {Gupta}, \citenamefont {Li},\ and\
  \citenamefont {Xiao}}]{PhysRevB.62.8931}%
  \BibitemOpen
  \bibfield  {author} {\bibinfo {author} {\bibfnamefont {L.}~\bibnamefont
  {Spinu}}, \bibinfo {author} {\bibfnamefont {H.}~\bibnamefont {Srikanth}},
  \bibinfo {author} {\bibfnamefont {A.}~\bibnamefont {Gupta}}, \bibinfo
  {author} {\bibfnamefont {X.~W.}\ \bibnamefont {Li}}, \ and\ \bibinfo {author}
  {\bibfnamefont {G.}~\bibnamefont {Xiao}},\ }\href {\doibase
  10.1103/PhysRevB.62.8931} {\bibfield  {journal} {\bibinfo  {journal} {Phys.
  Rev. B}\ }\textbf {\bibinfo {volume} {62}},\ \bibinfo {pages} {8931}
  (\bibinfo {year} {2000})}\BibitemShut {NoStop}%
\bibitem [{\citenamefont {Spinu}\ \emph {et~al.}(2003)\citenamefont {Spinu},
  \citenamefont {Stancu}, \citenamefont {Kubota}, \citenamefont {Ju},\ and\
  \citenamefont {Weller}}]{PhysRevB.68.220401}%
  \BibitemOpen
  \bibfield  {author} {\bibinfo {author} {\bibfnamefont {L.}~\bibnamefont
  {Spinu}}, \bibinfo {author} {\bibfnamefont {A.}~\bibnamefont {Stancu}},
  \bibinfo {author} {\bibfnamefont {Y.}~\bibnamefont {Kubota}}, \bibinfo
  {author} {\bibfnamefont {G.}~\bibnamefont {Ju}}, \ and\ \bibinfo {author}
  {\bibfnamefont {D.}~\bibnamefont {Weller}},\ }\href {\doibase
  10.1103/PhysRevB.68.220401} {\bibfield  {journal} {\bibinfo  {journal} {Phys.
  Rev. B}\ }\textbf {\bibinfo {volume} {68}},\ \bibinfo {pages} {220401}
  (\bibinfo {year} {2003})}\BibitemShut {NoStop}%
\bibitem [{\citenamefont {Rodriguez-Suarez}\ \emph {et~al.}(2003)\citenamefont
  {Rodriguez-Suarez}, \citenamefont {Vilela~Leao}, \citenamefont {de~Aguiar},
  \citenamefont {Rezende},\ and\ \citenamefont {Azevedo}}]{Rezende2003}%
  \BibitemOpen
  \bibfield  {author} {\bibinfo {author} {\bibfnamefont {R.~L.}\ \bibnamefont
  {Rodriguez-Suarez}}, \bibinfo {author} {\bibfnamefont {L.~H.}\ \bibnamefont
  {Vilela~Leao}}, \bibinfo {author} {\bibfnamefont {F.~M.}\ \bibnamefont
  {de~Aguiar}}, \bibinfo {author} {\bibfnamefont {S.~M.}\ \bibnamefont
  {Rezende}}, \ and\ \bibinfo {author} {\bibfnamefont {A.}~\bibnamefont
  {Azevedo}},\ }\href {\doibase 10.1063/1.1605816} {\bibfield  {journal}
  {\bibinfo  {journal} {Journal of Applied Physics}\ }\textbf {\bibinfo
  {volume} {94}},\ \bibinfo {pages} {4544} (\bibinfo {year}
  {2003})}\BibitemShut {NoStop}%
\bibitem [{\citenamefont {Cimpoesu}\ \emph {et~al.}(2007)\citenamefont
  {Cimpoesu}, \citenamefont {Stancu},\ and\ \citenamefont
  {Spinu}}]{PhysRevB.76.054409}%
  \BibitemOpen
  \bibfield  {author} {\bibinfo {author} {\bibfnamefont {D.}~\bibnamefont
  {Cimpoesu}}, \bibinfo {author} {\bibfnamefont {A.}~\bibnamefont {Stancu}}, \
  and\ \bibinfo {author} {\bibfnamefont {L.}~\bibnamefont {Spinu}},\ }\href
  {\doibase 10.1103/PhysRevB.76.054409} {\bibfield  {journal} {\bibinfo
  {journal} {Phys. Rev. B}\ }\textbf {\bibinfo {volume} {76}},\ \bibinfo
  {pages} {054409} (\bibinfo {year} {2007})}\BibitemShut {NoStop}%
\bibitem [{\citenamefont {Qiu}\ and\ \citenamefont {Bader}(2000)}]{Qiu2000}%
  \BibitemOpen
  \bibfield  {author} {\bibinfo {author} {\bibfnamefont {Z.~Q.}\ \bibnamefont
  {Qiu}}\ and\ \bibinfo {author} {\bibfnamefont {S.~D.}\ \bibnamefont
  {Bader}},\ }\href {\doibase 10.1063/1.1150496} {\bibfield  {journal}
  {\bibinfo  {journal} {Review of Scientific Instruments}\ }\textbf {\bibinfo
  {volume} {71}},\ \bibinfo {pages} {1243} (\bibinfo {year}
  {2000})}\BibitemShut {NoStop}%
\bibitem [{\citenamefont {Soldatov}\ and\ \citenamefont
  {Schäfer}(2017{\natexlab{a}})}]{doi:10.1063/1.4991820}%
  \BibitemOpen
  \bibfield  {author} {\bibinfo {author} {\bibfnamefont {I.~V.}\ \bibnamefont
  {Soldatov}}\ and\ \bibinfo {author} {\bibfnamefont {R.}~\bibnamefont
  {Schäfer}},\ }\href {\doibase 10.1063/1.4991820} {\bibfield  {journal}
  {\bibinfo  {journal} {Review of Scientific Instruments}\ }\textbf {\bibinfo
  {volume} {88}},\ \bibinfo {pages} {073701} (\bibinfo {year}
  {2017}{\natexlab{a}})}\BibitemShut {NoStop}%
\bibitem [{\citenamefont {Soldatov}\ and\ \citenamefont
  {Schäfer}(2017{\natexlab{b}})}]{doi:10.1063/1.5003719}%
  \BibitemOpen
  \bibfield  {author} {\bibinfo {author} {\bibfnamefont {I.~V.}\ \bibnamefont
  {Soldatov}}\ and\ \bibinfo {author} {\bibfnamefont {R.}~\bibnamefont
  {Schäfer}},\ }\href {\doibase 10.1063/1.5003719} {\bibfield  {journal}
  {\bibinfo  {journal} {Journal of Applied Physics}\ }\textbf {\bibinfo
  {volume} {122}},\ \bibinfo {pages} {153906} (\bibinfo {year}
  {2017}{\natexlab{b}})}\BibitemShut {NoStop}%
\bibitem [{\citenamefont {McCord}\ \emph {et~al.}(2003)\citenamefont {McCord},
  \citenamefont {Sch{\"a}fer}, \citenamefont {Mattheis},\ and\ \citenamefont
  {Barholz}}]{mccord2003kerr}%
  \BibitemOpen
  \bibfield  {author} {\bibinfo {author} {\bibfnamefont {J.}~\bibnamefont
  {McCord}}, \bibinfo {author} {\bibfnamefont {R.}~\bibnamefont {Sch{\"a}fer}},
  \bibinfo {author} {\bibfnamefont {R.}~\bibnamefont {Mattheis}}, \ and\
  \bibinfo {author} {\bibfnamefont {K.-U.}\ \bibnamefont {Barholz}},\
  }\href@noop {} {\bibfield  {journal} {\bibinfo  {journal} {Journal of applied
  physics}\ }\textbf {\bibinfo {volume} {93}},\ \bibinfo {pages} {5491}
  (\bibinfo {year} {2003})}\BibitemShut {NoStop}%
\bibitem [{\citenamefont {Gornakov}\ \emph {et~al.}(2006)\citenamefont
  {Gornakov}, \citenamefont {Kabanov}, \citenamefont {Tikhomirov},
  \citenamefont {Nikitenko}, \citenamefont {Urazhdin}, \citenamefont {Yang},
  \citenamefont {Chien}, \citenamefont {Shapiro},\ and\ \citenamefont
  {Shull}}]{PhysRevB.73.184428}%
  \BibitemOpen
  \bibfield  {author} {\bibinfo {author} {\bibfnamefont {V.~S.}\ \bibnamefont
  {Gornakov}}, \bibinfo {author} {\bibfnamefont {Y.~P.}\ \bibnamefont
  {Kabanov}}, \bibinfo {author} {\bibfnamefont {O.~A.}\ \bibnamefont
  {Tikhomirov}}, \bibinfo {author} {\bibfnamefont {V.~I.}\ \bibnamefont
  {Nikitenko}}, \bibinfo {author} {\bibfnamefont {S.~V.}\ \bibnamefont
  {Urazhdin}}, \bibinfo {author} {\bibfnamefont {F.~Y.}\ \bibnamefont {Yang}},
  \bibinfo {author} {\bibfnamefont {C.~L.}\ \bibnamefont {Chien}}, \bibinfo
  {author} {\bibfnamefont {A.~J.}\ \bibnamefont {Shapiro}}, \ and\ \bibinfo
  {author} {\bibfnamefont {R.~D.}\ \bibnamefont {Shull}},\ }\href {\doibase
  10.1103/PhysRevB.73.184428} {\bibfield  {journal} {\bibinfo  {journal} {Phys.
  Rev. B}\ }\textbf {\bibinfo {volume} {73}},\ \bibinfo {pages} {184428}
  (\bibinfo {year} {2006})}\BibitemShut {NoStop}%
\bibitem [{\citenamefont {Vansteenkiste}\ \emph {et~al.}(2014)\citenamefont
  {Vansteenkiste}, \citenamefont {Leliaert}, \citenamefont {Dvornik},
  \citenamefont {Helsen}, \citenamefont {Garcia-Sanchez},\ and\ \citenamefont
  {{Van Waeyenberge}}}]{Vansteenkiste2014}%
  \BibitemOpen
  \bibfield  {author} {\bibinfo {author} {\bibfnamefont {A.}~\bibnamefont
  {Vansteenkiste}}, \bibinfo {author} {\bibfnamefont {J.}~\bibnamefont
  {Leliaert}}, \bibinfo {author} {\bibfnamefont {M.}~\bibnamefont {Dvornik}},
  \bibinfo {author} {\bibfnamefont {M.}~\bibnamefont {Helsen}}, \bibinfo
  {author} {\bibfnamefont {F.}~\bibnamefont {Garcia-Sanchez}}, \ and\ \bibinfo
  {author} {\bibfnamefont {B.}~\bibnamefont {{Van Waeyenberge}}},\ }\href
  {\doibase 10.1063/1.4899186} {\bibfield  {journal} {\bibinfo  {journal} {AIP
  Advances}\ }\textbf {\bibinfo {volume} {4}},\ \bibinfo {pages} {107133}
  (\bibinfo {year} {2014})}\BibitemShut {NoStop}%
\bibitem [{\citenamefont {Leliaert}\ \emph {et~al.}(2014)\citenamefont
  {Leliaert}, \citenamefont {Van~de Wiele}, \citenamefont {Vansteenkiste},
  \citenamefont {Laurson}, \citenamefont {Durin}, \citenamefont {Dupr{\'e}},\
  and\ \citenamefont {Van~Waeyenberge}}]{Lel2014}%
  \BibitemOpen
  \bibfield  {author} {\bibinfo {author} {\bibfnamefont {J.}~\bibnamefont
  {Leliaert}}, \bibinfo {author} {\bibfnamefont {B.}~\bibnamefont {Van~de
  Wiele}}, \bibinfo {author} {\bibfnamefont {A.}~\bibnamefont {Vansteenkiste}},
  \bibinfo {author} {\bibfnamefont {L.}~\bibnamefont {Laurson}}, \bibinfo
  {author} {\bibfnamefont {G.}~\bibnamefont {Durin}}, \bibinfo {author}
  {\bibfnamefont {L.}~\bibnamefont {Dupr{\'e}}}, \ and\ \bibinfo {author}
  {\bibfnamefont {B.}~\bibnamefont {Van~Waeyenberge}},\ }\href {\doibase
  10.1063/1.4883297} {\bibfield  {journal} {\bibinfo  {journal} {Journal of
  Applied Physics}\ }\textbf {\bibinfo {volume} {115}},\ \bibinfo {pages}
  {233903} (\bibinfo {year} {2014})}\BibitemShut {NoStop}%
\bibitem [{\citenamefont {Exl}\ \emph {et~al.}(2014)\citenamefont {Exl},
  \citenamefont {Bance}, \citenamefont {Reichel}, \citenamefont {Schrefl},
  \citenamefont {{Peter Stimming}},\ and\ \citenamefont {Mauser}}]{Exl2014}%
  \BibitemOpen
  \bibfield  {author} {\bibinfo {author} {\bibfnamefont {L.}~\bibnamefont
  {Exl}}, \bibinfo {author} {\bibfnamefont {S.}~\bibnamefont {Bance}}, \bibinfo
  {author} {\bibfnamefont {F.}~\bibnamefont {Reichel}}, \bibinfo {author}
  {\bibfnamefont {T.}~\bibnamefont {Schrefl}}, \bibinfo {author} {\bibfnamefont
  {H.}~\bibnamefont {{Peter Stimming}}}, \ and\ \bibinfo {author}
  {\bibfnamefont {N.~J.}\ \bibnamefont {Mauser}},\ }\href {\doibase
  10.1063/1.4862839} {\bibfield  {journal} {\bibinfo  {journal} {Journal of
  Applied Physics}\ }\textbf {\bibinfo {volume} {115}},\ \bibinfo {pages}
  {17D118} (\bibinfo {year} {2014})}\BibitemShut {NoStop}%
\bibitem [{\citenamefont {Fulcomer}\ and\ \citenamefont
  {Charap}(1972)}]{Fulcomer1972}%
  \BibitemOpen
  \bibfield  {author} {\bibinfo {author} {\bibfnamefont {E.}~\bibnamefont
  {Fulcomer}}\ and\ \bibinfo {author} {\bibfnamefont {S.~H.}\ \bibnamefont
  {Charap}},\ }\href@noop {} {\bibfield  {journal} {\bibinfo  {journal}
  {Journal of Applied Physics}\ }\textbf {\bibinfo {volume} {43}},\ \bibinfo
  {pages} {4190} (\bibinfo {year} {1972})}\BibitemShut {NoStop}%
\bibitem [{\citenamefont {Stiles}\ and\ \citenamefont
  {McMichael}(1999)}]{PhysRevB.59.3722}%
  \BibitemOpen
  \bibfield  {author} {\bibinfo {author} {\bibfnamefont {M.~D.}\ \bibnamefont
  {Stiles}}\ and\ \bibinfo {author} {\bibfnamefont {R.~D.}\ \bibnamefont
  {McMichael}},\ }\href {\doibase 10.1103/PhysRevB.59.3722} {\bibfield
  {journal} {\bibinfo  {journal} {Phys. Rev. B}\ }\textbf {\bibinfo {volume}
  {59}},\ \bibinfo {pages} {3722} (\bibinfo {year} {1999})}\BibitemShut
  {NoStop}%
\bibitem [{\citenamefont {O’Grady}\ \emph {et~al.}(2010)\citenamefont
  {O’Grady}, \citenamefont {Fernandez-Outon},\ and\ \citenamefont
  {Vallejo-Fernandez}}]{OGRADY2010883}%
  \BibitemOpen
  \bibfield  {author} {\bibinfo {author} {\bibfnamefont {K.}~\bibnamefont
  {O’Grady}}, \bibinfo {author} {\bibfnamefont {L.}~\bibnamefont
  {Fernandez-Outon}}, \ and\ \bibinfo {author} {\bibfnamefont {G.}~\bibnamefont
  {Vallejo-Fernandez}},\ }\href {\doibase
  https://doi.org/10.1016/j.jmmm.2009.12.011} {\bibfield  {journal} {\bibinfo
  {journal} {Journal of Magnetism and Magnetic Materials}\ }\textbf {\bibinfo
  {volume} {322}},\ \bibinfo {pages} {883} (\bibinfo {year}
  {2010})}\BibitemShut {NoStop}%
\bibitem [{\citenamefont {He}\ \emph {et~al.}(2010)\citenamefont {He},
  \citenamefont {Wang}, \citenamefont {Wu}, \citenamefont {Caruso},
  \citenamefont {Vescovo}, \citenamefont {Belashchenko}, \citenamefont
  {Dowben},\ and\ \citenamefont {Binek}}]{He2010}%
  \BibitemOpen
  \bibfield  {author} {\bibinfo {author} {\bibfnamefont {X.}~\bibnamefont
  {He}}, \bibinfo {author} {\bibfnamefont {Y.}~\bibnamefont {Wang}}, \bibinfo
  {author} {\bibfnamefont {N.}~\bibnamefont {Wu}}, \bibinfo {author}
  {\bibfnamefont {A.~N.}\ \bibnamefont {Caruso}}, \bibinfo {author}
  {\bibfnamefont {E.}~\bibnamefont {Vescovo}}, \bibinfo {author} {\bibfnamefont
  {K.~D.}\ \bibnamefont {Belashchenko}}, \bibinfo {author} {\bibfnamefont
  {P.~A.}\ \bibnamefont {Dowben}}, \ and\ \bibinfo {author} {\bibfnamefont
  {C.}~\bibnamefont {Binek}},\ }\href {\doibase 10.1038/nmat2785} {\bibfield
  {journal} {\bibinfo  {journal} {Nature Materials}\ }\textbf {\bibinfo
  {volume} {9}},\ \bibinfo {pages} {579} (\bibinfo {year} {2010})}\BibitemShut
  {NoStop}%
\bibitem [{\citenamefont {Wu}\ \emph {et~al.}(2015)\citenamefont {Wu},
  \citenamefont {Miao}, \citenamefont {Xu}, \citenamefont {Yan}, \citenamefont
  {Reeve}, \citenamefont {Zhang},\ and\ \citenamefont {Jiang}}]{Wu2015}%
  \BibitemOpen
  \bibfield  {author} {\bibinfo {author} {\bibfnamefont {S.~Z.}\ \bibnamefont
  {Wu}}, \bibinfo {author} {\bibfnamefont {J.}~\bibnamefont {Miao}}, \bibinfo
  {author} {\bibfnamefont {X.~G.}\ \bibnamefont {Xu}}, \bibinfo {author}
  {\bibfnamefont {W.}~\bibnamefont {Yan}}, \bibinfo {author} {\bibfnamefont
  {R.}~\bibnamefont {Reeve}}, \bibinfo {author} {\bibfnamefont {X.~H.}\
  \bibnamefont {Zhang}}, \ and\ \bibinfo {author} {\bibfnamefont
  {Y.}~\bibnamefont {Jiang}},\ }\href@noop {} {\bibfield  {journal} {\bibinfo
  {journal} {Scientific Reports}\ }\textbf {\bibinfo {volume} {5}} (\bibinfo
  {year} {2015})}\BibitemShut {NoStop}%
\end{thebibliography}%
\bibliographystyle{apsrev4-1}
\end{document}